\documentclass[twocolumn,trackchanges]{aastex62}

\submitjournal{ApJ}
\usepackage{amsmath}
\usepackage{comment}

\shorttitle{Unveiling the Role of Strong H$\alpha$-emitters during the Epoch of Reionization}
\shortauthors{Rinaldi et al.}

\begin{document}

%\title{MIDIS: The Role of Strong H$\alpha$ Emitters at $z\simeq7-8$ during the Epoch of Reionization - Constraints from JWST}

\title{\bf {MIDIS:  Unveiling the Role of Strong H$\alpha$-emitters during the Epoch of Reionization with \textit{JWST}}}

\newcommand{\gsim}{{\;\raise0.3ex\hbox{$>$\kern-0.75em\raise-1.1ex\hbox{$\sim$}}\;}}

\correspondingauthor{Pierluigi Rinaldi}
\email{rinaldi@astro.rug.nl}

\author[0000-0002-5104-8245]{P. Rinaldi}
\affiliation{Kapteyn Astronomical Institute, University of Groningen,
P.O. Box 800, 9700AV Groningen,
The Netherlands
}

\author[0000-0001-8183-1460]{K. I. Caputi}
\affiliation{Kapteyn Astronomical Institute, University of Groningen,
P.O. Box 800, 9700AV Groningen,
The Netherlands
}
\affiliation{Cosmic Dawn Center (DAWN), Copenhagen, Denmark
}

\author[0000-0000-0000-0000]{E. Iani}
\affiliation{Kapteyn Astronomical Institute, University of Groningen,
P.O. Box 800, 9700AV Groningen,
The Netherlands
}

\author[0000-0001-6820-0015]{L. Costantin}
\affiliation{Centro de Astrobiolog\'{\i}a (CAB), CSIC-INTA, Ctra. de Ajalvir km 4, Torrej\'on de Ardoz, E-28850, Madrid, Spain}

\author[0000-0001-9885-4589]{S. Gillman}
\affiliation{Cosmic Dawn Center (DAWN), Denmark}

\affiliation{DTU-Space, Elektrovej, Building 328 , 2800, Kgs. Lyngby, Denmark}

\author[0000-0003-4528-5639]{P. G. P\'erez-Gonz\'alez}
\affiliation{Centro de Astrobiolog\'{\i}a (CAB), CSIC-INTA, Ctra. de Ajalvir km 4, Torrej\'on de Ardoz, E-28850, Madrid, Spain}

\author[0000-0002-3005-1349]{G. \"Ostlin}
\affiliation{Department of Astronomy, Stockholm University, Oscar Klein Centre, AlbaNova University Centre, 106 91 Stockholm, Sweden}

\author[0000-0002-9090-4227]{L. Colina}
\affiliation{Centro de Astrobiolog\'{\i}a (CAB), CSIC-INTA, Ctra. de Ajalvir km 4, Torrej\'on de Ardoz, E-28850, Madrid, Spain}
\affiliation{Cosmic Dawn Center (DAWN), Denmark}

\author[0000-0002-2554-1837]{T. R. Greve}
\affiliation{Cosmic Dawn Center (DAWN), Denmark}
\affiliation{DTU-Space, Elektrovej, Building 328 , 2800, Kgs. Lyngby, Denmark}

\author[0000-0000-0000-0000]{H. U. N\o{}rgard-Nielsen}
\affiliation{Cosmic Dawn Center (DAWN), Denmark}
\affiliation{DTU-Space, Elektrovej, Building 328 , 2800, Kgs. Lyngby, Denmark}

\author[0000-0000-0000-0000]{G. S. Wright}
\affiliation{UK Astronomy Technology Centre, Royal Observatory Edinburgh,
Blackford Hill, Edinburgh EH9 3HJ, UK}

\author[0000-0002-7093-1877]{J. \'Alvarez-M\'arquez}
\affiliation{Centro de Astrobiolog\'{\i}a (CAB), CSIC-INTA, Ctra. de Ajalvir km 4, Torrej\'on de Ardoz, E-28850, Madrid, Spain}

\author[0000-0000-0000-0000]{A. Eckart}
\affiliation{I.Physikalisches Institut der Universit\"at zu K\"oln, Z\"ulpicher Str. 77,
50937 K\"oln, Germany}

\author[0000-0000-0000-0000]{M. Garc\'{\i}a-Mar\'{\i}n}
\affiliation{European Space Agency/Space Telescope Science Institute, 3700 San Martin Drive, Baltimore MD 21218, USA}

\author[0000-0002-4571-2306]{J. Hjorth}
\affiliation{DARK, Niels Bohr Institute, University of Copenhagen, Jagtvej 128,
2200 Copenhagen, Denmark}

\author[0000-0002-7303-4397]{O. Ilbert}
\affiliation{Aix Marseille Universit\'e, CNRS, LAM (Laboratoire d’Astrophysique de Marseille) UMR 7326, 13388, Marseille, France}

\author[0000-0002-7612-0469]{S. Kendrew}
\affiliation{European Space Agency/Space Telescope Science Institute, 3700 San Martin Drive, Baltimore MD 21218, USA}

\author[0000-0002-0690-8824]{A. Labiano}
\affiliation{Telespazio UK for the European Space Agency (ESA), ESAC, Camino Bajo del Castillo s/n, 28692 Villanueva de la Ca\~nada, Spain}

\author[0000-0000-0000-0000]{O. Le F\`evre}
\affiliation{Aix Marseille Universit\'e, CNRS, LAM (Laboratoire d’Astrophysique de Marseille) UMR 7326, 13388, Marseille, France}

\author[0000-0002-0932-4330]{J. Pye}
\affiliation{School of Physics \& Astronomy, Space Research Centre, Space Park Leicester, University of Leicester, 92 Corporation Road, Leicester, LE4 5SP, UK}

\author[0000-0000-0000-0000]{T. Tikkanen}
\affiliation{School of Physics \& Astronomy, Space Research Centre, Space Park Leicester, University of Leicester, 92 Corporation Road, Leicester, LE4 5SP, UK}

\author[0000-0003-4793-7880]{F. Walter}
\affiliation{Max-Planck-Institut f\"ur Astronomie, K\"onigstuhl 17, 69117 Heidelberg, Germany}

\author[00000-0001-5434-5942]{P. van der Werf}
\affiliation{Leiden Observatory, Leiden University, PO Box 9513, 2300 RA Leiden, The Netherlands}

\author[0000-0003-1810-0889]{M. Ward}
\affiliation{Centre for Extragalactic Astronomy, Durham University, South
Road, Durham DH1 3LE, UK}

\author[0000-0002-8053-8040]{M. Annunziatella}
\affiliation{Centro de Astrobiolog\'{\i}a (CAB), CSIC-INTA, Ctra. de Ajalvir km 4, Torrej\'on de Ardoz, E-28850, Madrid, Spain}
\affiliation{INAF-Osservatorio Astronomico di Capodimonte, Via Moiariello 16, I-80131 Napoli, Italy}

\author[0000-0002-0438-0886]{R. Azzollini}
\affiliation{Centro de Astrobiolog\'{\i}a (CAB), CSIC-INTA, Ctra. de Ajalvir km 4, Torrej\'on de Ardoz, E-28850, Madrid, Spain}
\affiliation{Dublin Institute for Advanced Studies, Astronomy \& Astrophysics Section, 31 Fitzwilliam Place, Dublin 2, Ireland}

\author[0000-0001-8068-0891]{A. Bik}
\affiliation{Department of Astronomy, Stockholm University, Oscar Klein Centre, AlbaNova University Centre, 106 91 Stockholm, Sweden}

\author[0000-0002-3952-8588]{L. Boogaard}
\affiliation{Max-Planck-Institut f\"ur Astronomie, K\"onigstuhl 17, 69117 Heidelberg, Germany}

\author[0000-0001-8582-7012]{S. E. I. Bosman}
\affiliation{Max-Planck-Institut f\"ur Astronomie, K\"onigstuhl 17, 69117 Heidelberg, Germany}

\author[0000-0003-2119-277X]{A. Crespo G\'{o}mez}
\affiliation{Centro de Astrobiolog\'{\i}a (CAB), CSIC-INTA, Ctra. de Ajalvir km 4, Torrej\'on de Ardoz, E-28850, Madrid, Spain}

\author[0000-0002-2624-1641]{I. Jermann}
\affiliation{Cosmic Dawn Center (DAWN), Denmark}
\affiliation{DTU-Space, Elektrovej, Building 328 , 2800, Kgs. Lyngby, Denmark}

\author[0000-0001-5710-8395]{D. Langeroodi}
\affiliation{DARK, Niels Bohr Institute, University of Copenhagen, Jagtvej 128, 2200 Copenhagen, Denmark}

\author[0000-0003-0470-8754]{J. Melinder}
\affiliation{Department of Astronomy, Stockholm University, Oscar Klein Centre, AlbaNova University Centre, 106 91 Stockholm, Sweden}

\author[0000-0001-5492-4522]{R. A. Meyer}
\affiliation{Max-Planck-Institut f\"ur Astronomie, K\"onigstuhl 17, 69117 Heidelberg, Germany}

\author[0000-0002-3305-9901]{T. Moutard}
\affiliation{Aix Marseille Universit\'e, CNRS, LAM (Laboratoire d’Astrophysique de Marseille) UMR 7326, 13388, Marseille,
France}

\author[0000-0000-0000-0000]{F. Peissker}
\affiliation{I.Physikalisches Institut der Universit\"at zu K\"oln, Z\"ulpicher Str. 77,
50937 K\"oln, Germany}

\author[0000-0001-7591-1907]{E. van Dishoeck}
\affiliation{Leiden Observatory, Leiden University, PO Box 9513, 2300 RA Leiden, The Netherlands}

\author[0000-0001-9818-0588]{M. G\"udel}
\affiliation{Dept. of Astrophysics, University of Vienna, Türkenschanzstr 17, A-1180 Vienna, Austria}
\affiliation{ETH Zürich, Institute for Particle Physics and Astrophysics, Wolfgang-Pauli-Str. 27, 8093 Zürich, Switzerland}

\author[0000-0002-1493-300X]{Th. Henning}
\affiliation{Max-Planck-Institut f\"ur Astronomie, K\"onigstuhl 17, 69117 Heidelberg, Germany}

\author[0000-0000-0000-0000]{P.-O. Lagage}
\affiliation{AIM, CEA, CNRS, Universit\'e Paris-Saclay, Universit\'e Paris
Diderot, Sorbonne Paris Cit\'e, F-91191 Gif-sur-Yvette, France}

\author[0000-0000-0000-0000]{T. Ray}
\affiliation{Dublin Institute for Advanced Studies, Astronomy \& Astrophysics Section, 31 Fitzwilliam Place, Dublin 2, Ireland}

\author[0000-0000-0000-0000]{B. Vandenbussche}
\affiliation{Institute of Astronomy, KU Leuven, Celestijnenlaan 200D bus 2401,
3001 Leuven, Belgium}

\author[0000-0000-0000-0000]{C. Waelkens}
\affiliation{Institute of Astronomy, KU Leuven, Celestijnenlaan 200D bus 2401,
3001 Leuven, Belgium}

\author[0000-0001-8460-1564]{Pratika Dayal}
\affiliation{Kapteyn Astronomical Institute, University of Groningen,
P.O. Box 800, 9700AV Groningen,
The Netherlands
}

\begin{abstract}

By using the ultra-deep \textit{JWST}/MIRI image at 5.6 $\mu m$ in the Hubble eXtreme Deep Field, we constrain the role of strong H$\alpha$-emitters (HAEs) during Cosmic Reionization at $z\simeq7-8$. Our sample of HAEs is comprised of young ($<35\;\rm Myr$) galaxies, except for one single galaxy ($\approx 300\;\rm Myr$), with low stellar masses ($\lesssim 10^{9}\;\rm M_{\odot}$). These HAEs show a wide range of UV-$\beta$ slopes, with a median value of $\beta = -2.15\pm0.21$ which broadly correlates with stellar mass. We estimate the ionizing photon production efficiency ($\xi_{ion,0}$) of these sources (assuming $f_{esc,LyC} = 0\%$), which yields a median value $\rm log_{10}(\xi_{ion,0}/(Hz\;erg^{-1})) = 25.50^{+0.10}_{-0.12}$. We show that $\xi_{ion,0}$ positively correlates with EW$_{0}$(H$\alpha$) and specific star formation rate (sSFR). Instead $\xi_{ion,0}$ weakly anti-correlates with stellar mass and $\beta$. Based on the $\beta$ values, we predict $f_{esc, LyC}=4\%^{+3}_{-2}$, which results in $\rm log_{10}(\xi_{ion}/(Hz\;erg^{-1})) = 25.55^{+0.11}_{-0.13}$. Considering this and related findings from the literature, we find a mild evolution of $\xi_{ion}$with redshift. Additionally, our results suggest that these HAEs require only modest escape fractions ($f_{esc, rel}$) of 6$-$15\% to reionize their surrounding intergalactic medium. By only considering the contribution of these HAEs, we estimated their total ionizing emissivity ($\dot{N}_{ion}$) as $\dot{N}_{ion} = 10^{50.53 \pm 0.45}; \text{s}^{-1}\text{Mpc}^{-3}$. When comparing their $\dot{N}_{ion}$ with “non-H$\alpha$ emitter” galaxies across the same redshift range, we find that that strong, young, and low-mass emitters may have played an important role during Cosmic Reionization.
\end{abstract}

.

\section{Introduction} \label{Introduction}
The Epoch of Reionization (EoR) represents one of the landmark events in the cosmic timeline. It refers to the last phase transition of hydrogen that occurred in the \textit{recent} Universe’s history, where the first generations of galaxies shaped it into the state we see it today \citep{Stiavelli_2009, Dayal_2018}. That moment refers to the period of cosmic history in which the neutral hydrogen in the intergalactic medium (IGM) had been reionized and had become transparent to Lyman continuum (LyC) radiation. \textit{How did the Universe reionize?} \textit{What drove the Cosmic Reionization?} Answering these questions is, nowadays, one of the key goals for modern astronomers. Theoretical predictions suggest that a combination of the first metal-free Population III stars  \citep[PopIII;][]{Bromm_2004}, the subsequent Population II stars, and mini-quasars and quasars can be pinpointed as the main culprits that reionized the Universe with their ultraviolet (UV) photons \citep[e.g.,][]{Venkatesan_2001}. These sources were believed to produce a sufficient amount of ionizing photons (E $\geq$ 13.6 eV) that could potentially escape the interstellar medium (ISM) and reionize the surrounding IGM. 

Over the last decades, star-forming galaxies have been proposed to be the preferred sources of ionizing photons \citep[e.g.,][]{Robertson_2010, Robertson_2015, Stark_2016, Dayal_2018, Jiang_2022, Robertson_2022, Trebitsch_2022, Matsuoka_2023} and many studies suggest that Cosmic Reionization ended, roughly speaking, 1 Gyr after the Big Bang \citep[$z\simeq5-6$; e.g.,][]{Lu_2022, Gaikwad_2023}. Nevertheless, understanding when Cosmic Reionization ended is still a matter of debate. Until last year, a vast amount of Lyman-break galaxies (LBGs) at $z>6$ had been identified from deep \textit{Hubble Space Telescope} (\textit{HST}) images \citep[e.g.,][]{Oesch_2018, Salmon_2020}, offering the opportunity to study the UV luminosity function (LF) at very high redshift \citep[e.g.,][]{Atek_2015, Livermore_2017}. Those studies showed a clear picture: the UV-faint sources (M$_{UV} > -18$ mag) dominated the galaxy number counts during the Epoch of Reionization. Therefore, characterizing their properties, over the past decades, became one of the most important goals in modern-day astronomy. Particularly, deep \textit{HST} observations showed that UV-faint galaxies were characterized by having very blue rest-UV continuum slopes ($\beta$), ranging from $-2.5 \lesssim \beta \lesssim -2$ \citep[e.g.,][]{Dunlop_2012, Finkelstein_2012b, Bouwens_2014, Bhatawdekar_2021}. These studies pointed out that galaxies at $z\gtrsim6$ are considerably bluer than those at $z\simeq2-3$, with UV slopes often having $\beta < -2$. Moreover, many theoretical and observational studies suggested that a not-negligible contribution of ionizing photons comes from galaxies with low stellar mass ($\rm M_{\star}<10^{9}\; M_{\odot}$) as well, although the exact amount of the ionizing photon budget and how it changes with redshift is still under debate \citep[e.g.,][]{Finkelstein_2019, Bera_2022, Dayal_2022, Mutch_2023}.

Demonstrating that star-forming galaxies were the main source of reionization during EoR requires understanding how many energetic UV photons were produced by young stars and what fraction of them ($f_{esc}$)\footnote{There are multiple definitions of the escape fraction in the literature. $f_{esc}$ refers to the fraction of intrinsic LyC photons that escape into the IGM. This definition is convenient to use in theoretical and simulation studies where the true number of LyC photons produced is known from the SFR and initial mass function, which is also called \textit{absolute escape fraction} ($f_{esc, abs}$). Another definition is the \textit{relative escape fraction} ($f_{esc,rel}$), referring to the fraction of LyC photons that escape the galaxy relative to the fraction of escaping non-ionizing photons at 1500\;{\AA};} \citep[e.g.,][]{Alavi_2020} capable of ionizing hydrogen outside galaxies escaping without interacting with clouds of dust and hydrogen within galaxies.

In the last fifteen years, many studies suggested that the average $f_{esc}$ needed to explain that galaxies were the main \textit{cosmic reionizers} was around $10-20$ per cent \citep[e.g.,][]{Ouchi_2009, Robertson_2013, Robertson_2015, Finkelstein_2019}. A key point, in that regard, is understanding how LyC photons escape into the IGM and, thus, reionize it. For that reason, studying LyC leakers is essential \citep[e.g.,][]{Chisholm_2022, Choustikov_2023, Mascia_2023}.

Distant galaxies (up to $z\simeq9$) are extremely efficient at producing ionizing photons. In particular, a key quantity that can be studied is the ionizing photon production efficiency ($\xi_{ion}$), which has been shown to increase as a function of redshift \citep[e.g.,][]{Bouwens_2016, Matthee_2017, Faisst_2019, Endsley_2021, Stefanon_2022} -- an increase of $\xi_{ion}$ would imply that galaxies do not need a high value of $f_{esc}$ to have been able to reionize the surrounding IGM.

Since at $z\gtrsim 6$ we cannot directly measure the LyC radiation due to the increasing absorption by neutral hydrogen in the IGM along the line of sight \citep[e.g.,][]{Inoue_2014}, we should instead rely on hydrogen recombination lines that offer indirect evidence of ionizing photons. The most important one is the Lyman-$\alpha$ emission line \citep{Osterbrock_1989}. However, observations, over the past decades, have shown that the number counts of galaxies emitting Lyman-$\alpha$, i.e. Lyman Alpha Emitters (LAEs), dramatically drop at $z\gtrsim 6$ because of its resonant nature \citep[e.g.,][]{Morales_2021}. Fortunately, we can rely on the second strongest hydrogen recombination line: the H$\alpha$ emission line \citep[e.g.,][]{Stefanon_2022}. Thankfully, \textit{JWST} \citep{Gardner_2023} nowadays offers us the opportunity to study more systematically the H$\alpha$ emission line in individual galaxies at high redshift ($z\gtrsim 7$) with \textit{HST}-like spatial resolution \citep{Rinaldi_2023, Javi_2023}.

As proposed by \citet{Leitherer_94}, when it is present, we can use H$\alpha$ is present in combination with UV continuum measurements to constrain $\xi_{ion}$ \citep[e.g.,][]{Bouwens_2016, Chisholm_2022, Stefanon_2022}. By definition, $\xi_{ion}$ strongly depends on the LyC escape fraction ($f_{esc, LyC}$). 
However, since our knowledge of the effective $f_{esc, LyC}$ is highly uncertain, it is usually considered that $\xi_{ion} = \xi_{ion,0}$, which implies that $f_{esc, LyC}$ is assumed to be zero. 

Finally, another key quantity to study EoR is the total ionizing emissivity ($\dot N_{ion}$; i.e. the comoving density of ionizing photons emitted into the IGM) which is usually parameterized as the product of: the galaxy UV luminosity density ($\rho_{UV}$), $\xi_{ion}$, and $f_{esc}$ \citep[e.g.,][]{Robertson_2013, Robertson_2015, Robertson_2022}. If we assume that galaxies produce the bulk of ionizing photons during reionization, $\dot N_{ion}$ can give us hints about the contribution of star-forming galaxies in reionizing the Universe, which, in turn, allows us to build up theoretical models to describe Cosmic Reionization \citep[e.g., ]{Manson_2019}.

In this work, we make use of a sample of bright H$\alpha$ emitter (HAE) galaxies at $z\simeq 7-8$ that has been detected in the Hubble eXtreme Deep Field (XDF) by using the deepest image of the Universe at 5.6 $\mu m$. By studying this sample of HAEs, we aim to infer their $\xi_{ion}$ and thus try to constrain the role they played during Cosmic Reionization.

The paper is organized as follows. In Section \ref{section2}, we briefly describe our sample of 12 HAEs, which was first presented in \citet{Rinaldi_2023}. In Section \ref{section3}, we present our results: for each source, we derive $\beta$,  M$_{UV}$, $\xi_{ion,0}$ and estimate $f_{esc, LyC}$, which in turn allows us to infer $\xi_{ion}$.  In Section \ref{section4}, we put our sources in context and analyze the impact of strong HAEs during the Epoch of Reionization. Finally, we summarize our findings in Section \ref{section5}.

Throughout this paper, we consider a cosmology with $H_{0} = 70\; \rm km\;s^{-1}\;Mpc^{-1}$, $\Omega_{M} = 0.3$, and $\Omega_{\Lambda} =0.7$. All magnitudes are total and refer to the AB system \citep{Oke_1983}. A \citet{Chabrier_2003} initial mass function (IMF) is assumed (0.1--100 M$_{\odot}$).

To propagate uncertainties in all the quantities presented, we employed Markov chain Monte Carlo simulations (MCMC) by considering 1000 iterations each time and a general distribution (with skewness) to take into account asymmetrical error bars if they are present. 

\section{Datasets and Sample Selection} \label{section2}
In this Section, we present how we selected our sample of HAEs. We refer the reader to \citet{Rinaldi_2023} for a more detailed discussion. Here we briefly summarize what we have done in the previous paper. 

The Hubble eXtreme Deep Field \citep[XDF;][]{illingworth2013}, with its groundbreaking \textit{HST} observations, has been a crucial window into studying the early Universe for over 30 years. With the arrival of \textit{JWST}, we are now expanding these observations into the near- and mid-infrared, thanks to the Near Infrared Camera \citep[NIRCam;][]{Rieke_2005} and Mid-Infrared Instrument \citep[MIRI;][]{Rieke_2015}. We collected ancillary data from \textit{HST} in 13 bands ($0.2-1.6\; \mu m$). See \citet{Whitaker_2019} for more detailed information on these observations. Compared to \citet{Rinaldi_2023},  we enriched our data set in XDF by considering also public data from JADES NIRCam with medium and broadband. Below we list all the NIRCam programs adopted in this work: PID: 1180; PI:  Daniel Eisenstein, PID: 1210; PI: Nora Luetzgendorf, PID: 1895; PI: Pascal Oesch, and PID: 1963; PI: Christina C. Williams, Sandro Tacchella, and Michael Maseda \citep{Eisenstein_2023, Rieke_2023, Oesch_2023, Williams_2023}. Finally, we complemented both \textit{HST} and \textit{JWST}/NIRCam data with the MIRI 5.6 $\mu m$ imaging from the JWST Guaranteed Time
Observations (GTO) program: MIRI Deep Imaging Survey
(MIDIS; PID: 1283, PI: Göran stlin), which represents the deepest image of the Universe at these wavelengths \citep{Boogaard_2023, Iani_2023, Rinaldi_2023}.

We employed the software \textsc{SExtractor} \citep[][]{SExtractor} to detect the sources and measure their photometry in all the available filters  from the \textit{HST} and \textit{JWST}. We used \textsc{SExtractor} in dual-image mode by adopting a super-detection image that we created by combining photometric information from different bands. Once we created the catalog in XDF, we performed the spectral energy distribution (SED) fitting employing \textsc{LePHARE} \citep{LePhare_2011}. A full description of the adopted methodology for the photometry and SED fitting can be found in \citet[][Section 2.2 and 2.3 respectively]{Rinaldi_2023}.

We then focused on the redshift bin $z\simeq 7-8$ to look for (H$\beta$ + [OIII]) and H$\alpha$ emitters. We found 58 potential candidates.
By analyzing their flux excess in NIRCam/F430M, NIRCam/F444W, and MIRI/F560W, we found 18 candidates. Among them, 12 lie on the MIRI coverage and show an excess in MIRI/F560W that we identified as H$\alpha$ excess. A detailed explanation of how we selected these strong HAEs can be found in \citet[][Section 3]{Rinaldi_2023}. Finally, our sample of HAEs constitutes 20\% of the star-forming galaxies that we analyzed at $z\simeq 7-8$.

\section{Results}\label{section3}
\subsection{Measuring UV absolute magnitudes and UV-$\beta$ slopes}
Over the past decades, the UV continuum slope, the so-called UV-$\beta$ slope, has been adopted as a proxy to infer properties of galaxies at very high redshift such as age, metallicity, and dust \citep[e.g.,][]{Schaerer_2002, Bouwens_2010, Wilkins_2013, Chisholm_2022}. Many studies have commonly found that, at high redshifts ($z\gtrsim 6$), the UV-$\beta$ slope appears to be bluer than what we usually can retrieve at lower redshifts, reaching, on average, values of $\beta \simeq -2$ \citep[e.g.,][]{Dunlop_2012, Finkelstein_2012b, Bhatawdekar_2021}. In this section, we derive the absolute magnitude of UV ($M_{UV}$) and UV-$\beta$ slope for our sample of sources in $z\simeq7-8$, following the same prescription as presented in \citet{Castellano_2012}. Briefly, we adopt a power law ($F\propto \lambda^{\beta}$) for the UV spectral range. We estimate $\beta$ by fitting a linear relation through the observed magnitudes of each object:

\begin{equation}
m_{i} = -2.5\cdot(\beta + 2) \cdot \mathrm{log(\lambda_{eff,i})} + C,
\label{beta_eq}
\end{equation}

where $m_{i}$ refers to the observed magnitude of the i-th filter at its effective wavelength ($\lambda_{\rm eff,i}$). See Section 4 in \citet{Castellano_2012} for more details.

To estimate the UV-$\beta$ slope, we follow the same methodology as that presented in \citet{Iani_2023}. Thus, we consider the rest frame wavelength range $\rm \lambda \simeq 1300 - 2500$ {\AA}\; for our fit (i.e., the UV spectral range). For this purpose, we only consider filters that have a detection (i.e., we do not consider upper limits in our fit). Finally, we impose a minimum number of bands (i.e., 3 bands at least) for our fit.

Once we estimate the UV-$\beta$ slope values, we derive $M_{UV}$ at 1500\;\AA. For this purpose, we derive $M_{UV}$ at 1500\;\AA\ from the best fit of the UV continuum slope. 

\begin{figure}[ht!]
    \centering
    \includegraphics[width = 0.49 \textwidth, height = 0.30 \textheight]{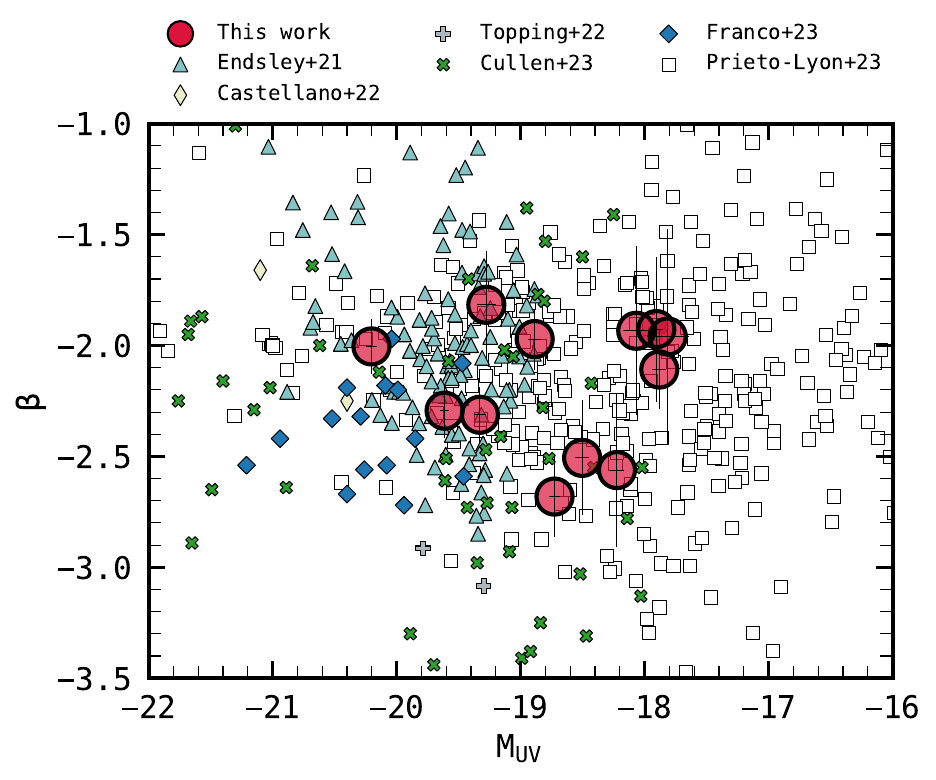}
    \caption{UV-$\beta$ slope as a function of the observed UV absolute magnitude. We compare our results with the recent literature at different redshifts \citep{Endsley_2021, Castellano_2022, Topping_2022, Cullen_2023, Franco_2023, Prieto_Lyon_2023}. We do not find any clear trend between $\beta$ and $M_{UV}$ at $z\simeq7-8$, although other studies claim it \citep[e.g., ][]{Cullen_2023}}.
    \label{beta_vs_MUV}
\end{figure}

In Figure \ref{beta_vs_MUV}, we show the relation between $\beta$ and $M_{UV}$ by considering our sample as well as the most recent literature at high redshift. We find that our sample has a median value of $\beta \simeq -2.15 \pm 0.21$ (16th and 84th percentile) which is in line with what has been found in the past at these redshifts ($z\simeq 7-8$) and  consistent, within the uncertainties, with the recent literature at high redshifts \citep[e.g.,][]{Endsley_2021, Cullen_2023}. In particular,  three of our galaxies have a very blue UV-$\beta$ slope ($-2.7 \leq \beta \leq -2.5$). Given their UV-$\beta$ slopes, they could be LyC leaker candidates with low metallicity \citep[e.g., ][]{Chisholm_2022}. Notwithstanding this, spectroscopic follow-up observations are needed to further investigate their nature.

Although the past literature has already shown that finding LyC leakers at $z > 5$ is challenging because the IGM transmission would not be high enough to observe the Lyman Continuum emission, it has been shown that the LyC leakage can be inferred, at such high redshifts, by using indirect indicators such as the UV-$\beta$ slope, Lyman-$\alpha$ emission line, absorption lines, and H$\beta$ \citep[e.g., ][]{Vanzella_2010, Leethochawalit_2016, Songaila_2018, Matthee_2018, Vanzella_2018, Bosman_2020,  Yamanaka_2020, Meyer_2021, Chisholm_2022, Begley_2023, Mascia_2023, Mascia_2023b, Roy_2023}

Such blue UV-$\beta$ slope values are not easily observed at intermediate redshifts ($z\simeq2-4$), and the candidates previously proposed at high redshifts, based on \textit{HST} data, were faint and had very uncertain values $\beta$. Instead, \textit{JWST}-based studies are now reporting more robust examples of sources with very blue UV-$\beta$ slopes at high redshifts \citep[e.g.,][]{Atek_2022, Castellano_2022, Topping_2022, Austin_2023, Bouwens_2023, Cullen_2023, Franco_2023, Saldana-Lopez_2023}.

In the last decade, a large number of studies have been conducted to study a possible relation between $\beta$ and $M_{UV}$, resulting in a debate that is still open at present. For instance, \citet{Dunlop_2012} reported that there is no correlation between $\beta$ and $M_{UV}$, although they only considered a sample of galaxies that had at least one 8$\sigma$ detection. Some other studies \citep[e.g.,][]{Bouwens_2012, Bouwens_2014}, instead, claimed that the UV continuum slopes of galaxies become bluer at fainter luminosities, although the dependence on redshift is still under discussion \citep[e.g.,][]{Cullen_2023}. We do not observe a clear correlation, neither with our own sample nor with the total data (our sample combined with the recent literature), but that this issue should be investigated further with larger samples. Other groups find a correlation  (\citealt{Topping_2022, Cullen_2023}) between these two quantities, but not all of them \citep[e.g., ][]{Dunlop_2012}. Therefore, this need to be investigated more in the future.

\begin{figure}[ht!]
    \centering
    \includegraphics[width = 0.49 \textwidth, height = 0.30 \textheight]{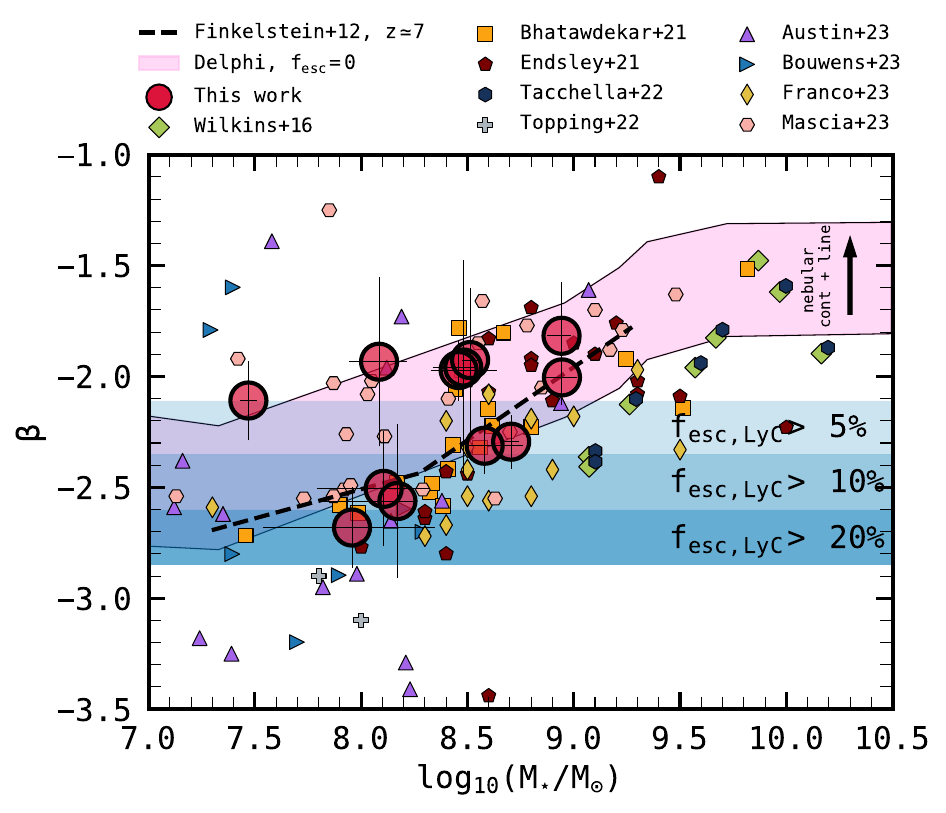}
    \caption{UV-$\beta$ slope as a function of stellar mass. A collection of results at high redshift from the recent literature is presented as well \citep{Wilkins_2016, Bhatawdekar_2021, Endsley_2021, Tacchella_2022, Topping_2022, Austin_2023, Bouwens_2023, Franco_2023, Mascia_2023}. From this plot, we can see that our sample of HAEs is dominated by low-mass galaxies ($\rm M_{\star} \leq 10^{9}\; M_{\odot}$). We also show colored regions (blue gradients) that correspond to the averages of the escape fraction of the Lyman continuum photons (5, 10, 20 per cent) by adopting Equation 11 from \citet{Chisholm_2022}. We include the $z \simeq 7$ relation from \cite{Finkelstein_2012b} as the dashed line. The purple shaded area refers to \textsc{Delphi} simulations at $z\simeq7$,  where we show how the nebular contribution (both continuum and emission lines) can impact the UV-$\beta$ slope as a function of $\rm M_{\star}$. Particularly, the lower limit of the shaded area refers to the pure stellar continuum + dust. The upper limit, instead, refers to the maximum contribution of stellar + nebular continuum + nebular lines + dust.}
    \label{Beta_vs_Mass}
\end{figure}

We also investigate if there is any correlation between $\beta$ and stellar mass (M$_{\star}$) -- see Figure \ref{Beta_vs_Mass}. The relation between these two quantities has been intensively studied at different redshifts \citep[e.g.,][]{Finkelstein_2012b} in the past years. In this work, we find that our galaxies span stellar masses $\rm log_{10} (M_{\star}/M_\odot) \simeq 7.5- 9$ at $z\simeq7-8$, similarly to most of the other recent studies at such high redshifts \citep[e.g.,][]{Topping_2022, Franco_2023}. We find that $\beta$ broadly correlates with M$_{\star}$, i.e. the most massive galaxies have flatter UV continua, following the relation proposed at $z\simeq 7$ in \citet{Finkelstein_2012b}. We also plot \textsc{Delphi} simulations, a semi-analytic model for early galaxy formation that couples the assembly of dark matter halos and their baryonic components \citep[][]{Dayal_2022, Mauerhofer_2023}. At $z\simeq7$, it can study the assembly of galaxies with stellar masses $\rm log_{10}(M_{\star}/M_{\odot}) = 6 - 12$. In addition to the key processes of mass assembly through both accretion and mergers, it has a dust model that has been fully calibrated against the latest ALMA results of the REBELS survey \citep{Bouwens_2022ALMA}. The beta slopes predicted by \textsc{Delphi} include the contribution from stellar and nebular emission (both from the continuum and emission lines) and the impact of dust attenuation as detailed in \citet{Mauerhofer_2023}. The UV dust attenuation, in \textsc{Delphi}, is convolved with a Calzetti extinction curve; in order to calculate the nebular emission, we use the escape fraction results from the \textit{Low-redshift Lyman Continuum Survey} \citep[LzLCS;][]{Chisholm_2022} as detailed in \citet{Trebitsch_2022}.

We also notice that $\beta$ becomes bluer at lower M$_{\star}$ as previously reported by \citet{Finkelstein_2012b, Bhatawdekar_2021} and recently suggested, at similar redshifts, in \citet{Franco_2023} by employing \textit{JWST} data. This relation can be explained by the fact that galaxies that are intensively forming stars, and, thus, are producing ionizing photons, rapidly synthesize metals and simultaneously grow in terms of stellar mass. Indeed, the more galaxies build up their stellar mass, the more they retain metals \citep[e.g.,][]{Tremonti_2004, Maiolino_2019} and, thus, create more dust \citep[e.g.,][]{Popping_2017, Mauerhofer_2023} which might explain why we find larger values of $\beta$ at higher stellar masses. In particular, in Figure \ref{Beta_vs_Mass}, we also display the expected Lyman Continuum escape fraction ($f_{esc, LyC}$) as shown in \citet[][blue shaded areas]{Chisholm_2022}. By looking at the expected $f_{esc, LyC}$, it appears that low-mass galaxies should be characterized by higher escape fraction values as predicted in many studies \citep[e.g.,][]{Dayal_2020, Trebitsch_2022}. In particular, \citet[][]{Mutch_2016} suggested that galaxies residing in halos of mass $M_{vir} \simeq 10^{8} - 10^{9}\;\rm M_{\odot}$ are dominant contributors of the ionizing budget of the Universe before Cosmic Reionization is complete. However, we warn the reader that the exact mass/magnitude range of sources that provide key reionization photons remains highly debated and model-dependent \citep[e.g., ][]{Dayal_2018}.

\begin{figure*}[ht!]
    \centering
    \includegraphics[width = 1 \textwidth]{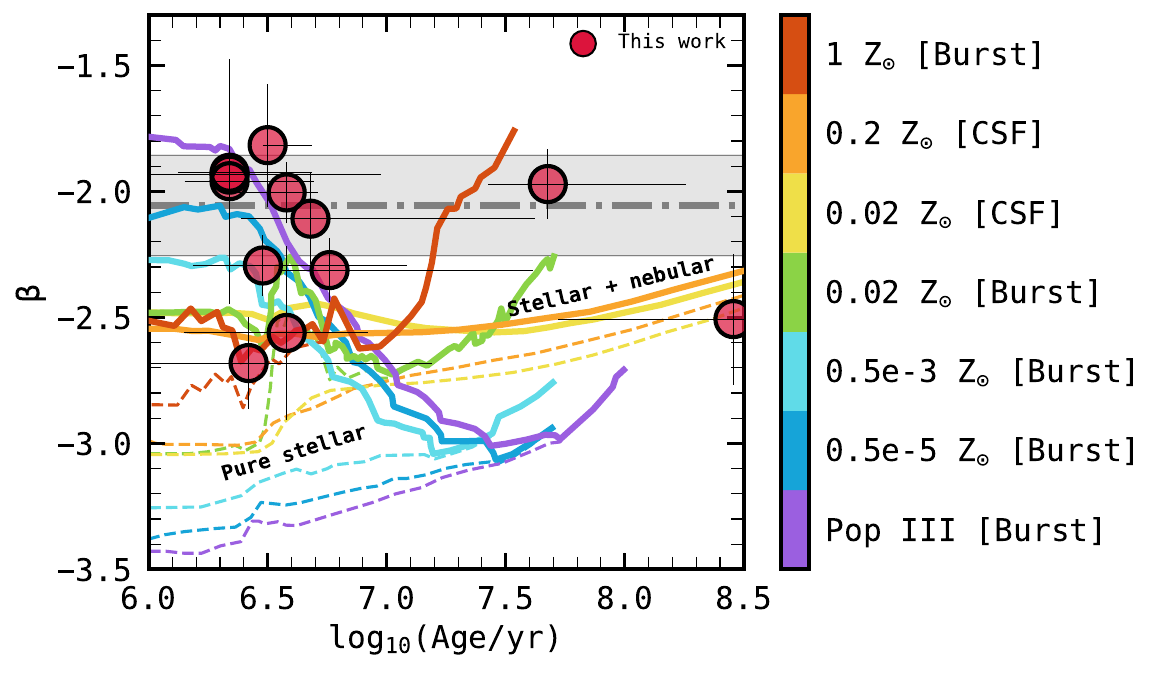}
    \caption{Beta slope as a function of galaxy age. The ages of galaxies have been obtained as output from \textsc{LePHARE}. The grey dashed line refers to the median $\beta$ value that we find in our sample, which is in line with what we expect from galaxies at high redshifts. For comparison, we also include theoretical predictions by considering synthetic-model tracks from \citet{Schaerer_2002, Schaerer_2003}, which are color-coded based on metallicity. Solid lines refer to models with a combination of stellar and nebular contributions, while dashed lines refer to pure stellar models. Two different SFHs have been adopted: burst and constant star formation.}
    \label{beta_vs_age_models}
\end{figure*}

In Figure \ref{beta_vs_age_models}, we show the behavior of $\beta$ as a function of the age for our galaxies, along with synthetic-model tracks from the literature \citep{Schaerer_2002, Schaerer_2003}, corresponding to different SFHs (burst and Constant Star Formation, hereafter CSF) and metallicities. In particular, the ages for our galaxies directly come from \textsc{LePHARE} and are purely based on the formation time as predicted by the \citet{BC_2003} models (i.e.,
the models we assumed to perform the SED fitting). We refer the reader to \citet{Rinaldi_2023} for more details regarding how the SED fitting has been performed. 
Here we show models that take into account pure stellar contribution (dashed lines) and stellar and nebular continuum emission (solid lines). We also show tracks that describe the expected trend for PopIII stars by considering only a single burst of star formation.

Our galaxies are all young (with ages $ \rm \lesssim 35\; Myr$), except for one single source that shows a stellar population a bit older compared to the rest of the sample ($\approx 300\;\rm Myr$), and, as discussed before, span $\beta$ values between $-2.7$ and $-1.8$, with a median $\beta \simeq -2.15 \pm 0.21$. Explaining this combination of parameters requires stellar models with nebular emission, as models with pure stellar contribution produce $\beta$ slopes which are significantly lower than our values. Our data points also suggest that our galaxies could span a range of metallicities, with some of them even being compatible with solar metallicity tracks. For some others, only very low metallicity values are possible ($\leq 0.02\;Z_{\odot}$).

\subsection{Inferring the ionizing photon production efficiency and the escape fraction of Lyman continuum photons}

In the past, numerous studies have demonstrated that detecting LyC radiation during the EoR is challenging at $z\gtrsim5-6$ due to the increasing optical depth along the line of sight \citep[][]{Inoue_2014, Fan_2022}. Interestingly, indirect evidence of ionizing photons can be retrieved from recombination lines because they are produced after photoionization has taken place. Observations have shown that the strongest among these lines is Lyman-$\alpha$ \citep{Osterbrock_1989}. However, many studies showed that the number counts of Lyman Alpha emitter (LAE) galaxies dramatically drop at $z\gtrsim6-7$ also because of the increasing neutral-hydrogen fraction in the IGM as a function of the redshift \citep[e.g.,][]{pentericci+14, Fuller_2020, Morales_2021}, although a few exceptional LAEs have been found at very high redshifts with \textit{JWST} \citep[e.g.,][]{Bunker_2023, Saxena_2023}.

Another option that we can rely on at $z\gtrsim6$ is the H$\alpha$ emission line which, unlike the Lyman-$\alpha$, is not affected by resonant scattering in the IGM. In particular, if we use the H$\alpha$ emission line in combination with a measure of the UV continuum, we can estimate the ionizing photon production efficiency. Interestingly, $\xi_{ion}$ indicates the connection between the observed rest-frame UV emission from galaxies and the corresponding amount of Lyman continuum photons emitted by their stars \citep[e.g.,][]{Nanayakkara_2020}. Therefore, this parameter is crucial to understanding the role of star-forming galaxies in the process of reionization because it gives an idea of the amount of the ionizing photons that they were actually able to produce in the early Universe \citep[e.g.,][]{Schaerer_2016}.  

In turn, the parameter $\xi_{ion}$ depends on the IMF, star formation histories (SFHs), the evolution of individual stars, and metallicity \citep[e.g.,][]{Shivaei_2018}. The value of $\xi_{ion}$ can be predicted from stellar-population synthesis models \citep[e.g.,][]{Eldridge_2022}. For instance, by analyzing \textsc{BlueTides} simulations, \citet{Wilkins_2016b} found that the choice of stellar population synthesis model (i.e., variations in SFHs and metal enrichment) for high-redshift galaxies can lead to $\mathrm{log_{10}(\xi_{ion}/(Hz\; erg^{-1}) \simeq  25.1 - 25.5}$, which is broadly consistent with recent observational constraints at high-redshift \citep[e.g.,][]{Stark_2015, Stark_2017, Endsley_2021, Sun_2022, Atek_2023, Bunker_2023, Whitler_2023}. The canonical value assumed for $\mathrm{log_{10}(\xi_{ion}/(Hz\; erg^{-1})}$ is  $25.2\pm 0.1$ \citep[e.g.,][]{Robertson_2013, Bouwens_2015}.  For instance, if we assume a constant star-formation history, $\xi_{ion}$ increases with metallicity and decreases with increasing $\beta$, saturating at $\beta \gsim -1.9$ \citep[][see their Figure 1]{Robertson_2013}.

\citet{Leitherer_94} have shown, by using an extensive grid of evolutionary synthesis models for populations of massive stars, that the H$\alpha$ luminosity ($L(H\alpha)$) from a galaxy is closely connected to its total Lyman-continuum luminosity.
Indeed,  following \citet{Leitherer_94}, we can define $\xi_{ion}$ as follows:

\begin{equation}
\xi_{ion} = \frac{L(H\alpha)}{(1-f_{esc, LyC})L_{UV, \nu}^{int}} \cdot 7.37 \times 10^{11}\rm \;Hz\;erg^{-1},
\label{Eq_2}
\end{equation}
where $L(H\alpha)$ refers to the intrinsic, i.e. unattenuated, luminosity in erg s$^{-1}$ and  $L_{UV, \nu}^{int}$ refers to intrinsic UV luminosity density in erg s$^{-1}$ Hz$^{-1}$ at 1500 {\AA}. 

We obtain the intrinsic $L(H\alpha)$ as we presented in \citet{Rinaldi_2023} by adopting the Calzetti reddening law \citep{Calzetti_2000}. To obtain $L_{UV, \nu}^{int}$, we employed the $\beta$ slope method \citep[e.g.,][]{Matthee_2017} as described in \citet{Meurer_99}. 

We know that $L_{UV, \nu}^{int} = $ $L_{UV,\nu}/f_{esc, UV}$, where $f_{esc, UV}$ is the fraction of emitted photons escaping their host galaxy in the UV continuum. Following the \citet{Meurer_99} prescription and employing \citet{Calzetti_2000}, we derive that:

\begin{equation}
f_{esc, UV} = 
\begin{cases}
10^{-0.83(2.23 + \beta)}, &  \text{$\beta$}\ > -2.23\\
1, & \text{otherwise}
\end{cases}
\end{equation}

In particular, $f_{esc, UV} = 1$ implies that the galaxies with a $\beta$ slope bluer than  $\beta< -2.23$ are assumed to be dust-free, so we do not correct for dust. 
Nevertheless, despite being an assumption in \citet{Meurer_99}, we caution the reader that a $\beta < -2.23$ does not necessarily imply the absence of dust extinction. Other parameters have been found to steepen the UV-$\beta$ slope to even bluer colors such as IMF, metallicity, and age \citep[e.g.,][]{Casey_2014, Cullen_2023, Franco_2023}.

Since our observations prevent us from directly calculating $f_{esc, LyC}$, here we assume that $f_{esc, LyC} = 0$. Therefore, by applying Eq. \ref{Eq_2}, we retrieve $\xi_{ion,0}$ ($\equiv \xi_{ion}$ when $f_{esc, LyC} = 0$).

We warn the reader that different methods can be used to estimate $f_{esc, UV}$ \citep[see][for more details]{Matthee_2017}. However, the recent literature has shown that estimating $f_{esc, UV}$ based on the UV-$\beta$ slope leads to $\xi_{ion}$ values more in line with what we expect at high redshift, which led us to adopt the same approach \citep[e.g.,][]{Matthee_2017, Shivaei_2018, Lam_2019, Prieto_Lyon_2023}.

\subsection{Comparison between $\xi_{ion,0}$ and stellar properties}
In the following subsections, we present the comparison between $\xi_{ion,0}$ and different stellar properties, such as $M_{UV}$, M$_{_{\star}}$, EW(H$\alpha$), etc., and compare our results with the literature at high redshifts, including the recent results with {\it JWST}. Remarkably, the correlations and anti-correlations between $\xi_{ion,0}$ and the stellar properties, that we are going to present in the following subsections, are very similar to what has been recently found at lower redshifts in \citet[][]{Castellano_2023} -- see their Figure 8.

Finally, we make use of the UV-$\beta$ slopes to predict the expected $f_{esc, LyC}$ for these HAEs by following the prescription presented in \citet{Chisholm_2022}.

\subsubsection{$\xi_{ion,0}$ versus UV Absolute Magnitude}

\begin{figure}[ht!]
    \centering
    \includegraphics[width = 0.49 \textwidth, height = 0.30 \textheight]{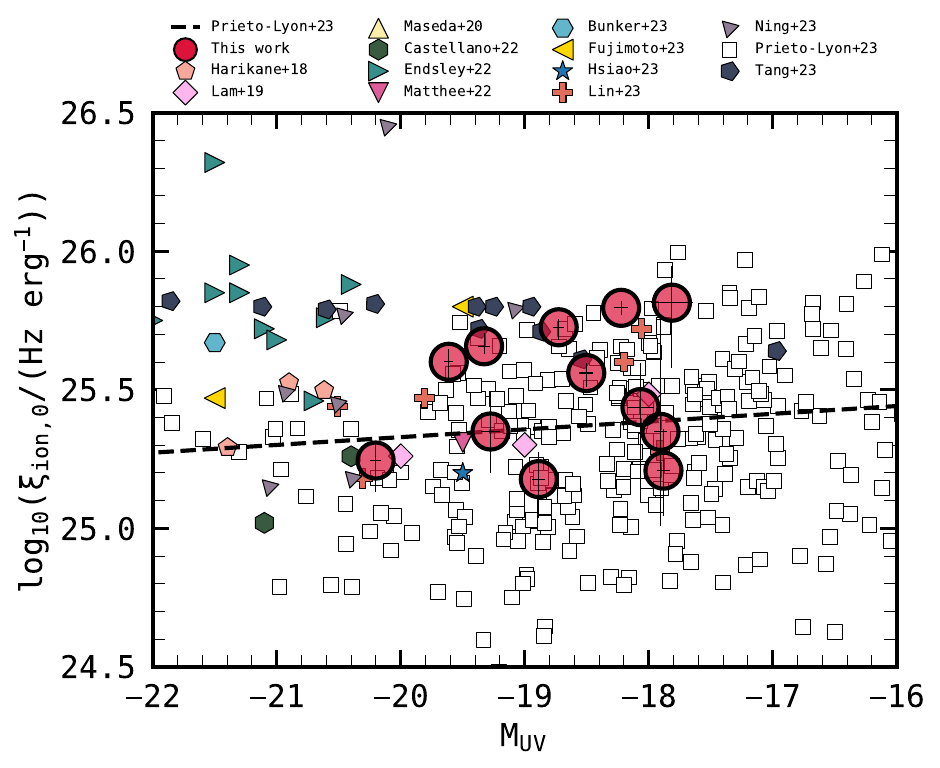}
    \caption{$\xi_{ion,0}$ as a function of M$_{UV}$. 
    We also collect data points from the recent literature at high redshifts \citep{Harikane_2018, Lam_2019, Maseda_2020, Castellano_2022, Endsley_2022, Matthee_2022, Bunker_2023, Fujimoto_2023, Hsiao_2023, Lin_2023, Ning_2023, Prieto_Lyon_2023, Tang_2023}. A weak correlation seems to be present between $\xi_{ion,0}$ and $M_{UV}$, in agreement with the recent literature \citep[e.g.,][]{Prieto_Lyon_2023, Simmonds_2023}. However, our sample is too small and future deep observations are needed to further constrain it.}
    \label{Xion_f_esc_MUV}    
\end{figure}

We know that $M_{UV}$ is one of the easiest quantities to measure for high-redshift galaxies. Particularly, the integral of the UV luminosity function is extremely important in determining the total ionizing emissivity of galaxies (see Section \ref{section4}). For that reason, we decided to investigate if there is a correlation between these two parameters. In Figure ~\ref{Xion_f_esc_MUV}, we show $\xi_{ion,0}$ versus  M$_{UV}$. We also compare our results with the most recent literature at high redshift. By looking at this plot, we find that our HAEs show a large variety of $\xi_{ion,0}$ values.  For our galaxy sample we find a median value of log$\mathrm{_{10}(\xi_{ion,0}/(Hz\;erg^{-1}))}$ $\simeq 25.49_{-0.12}^{+0.10}$ (16th and 84th percentile).
Although it is difficult to say, mostly due to our sample size, a weak correlation appears to be between $\xi_{ion,0}$ and M$_{UV}$, suggesting that faint galaxies, at high redshift, could be regarded as the bulk of ionizing photons that could potentially escape into the IGM and, thus, reionize it \citep{Duncan_2015}. A similar result has been found also in, e.g., \citet{Prieto_Lyon_2023, Simmonds_2023} by leveraging a bigger sample.

\subsubsection{$\xi_{ion,0}$ versus $\mathrm{EW_{0}(H\alpha)}$}
\begin{figure}[ht!]
    \centering
    \includegraphics[width = 0.49 \textwidth, height = 0.30 \textheight]{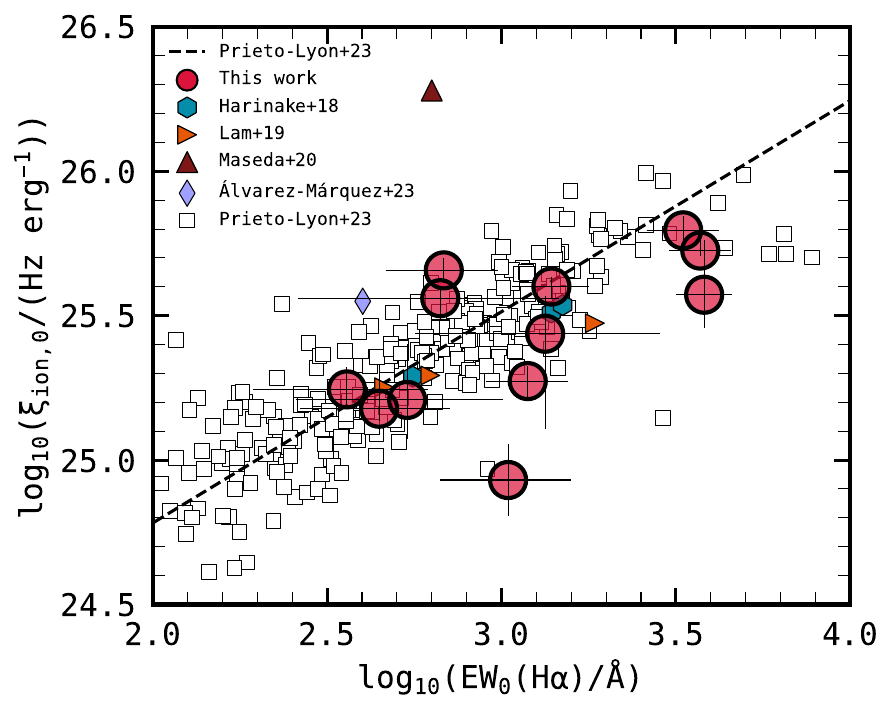}
    \caption{$\xi_{ion,0}$ as a function of $\rm EW_{0}(H\alpha)$. A collection of recent findings at high redshift is presented as well \citep{Harikane_2018, Lam_2019, Maseda_2020, Javi_2023, Prieto_Lyon_2023}. A correlation between these two quantities is evident, which agrees with the recent findings at lower redshifts \citep{Prieto_Lyon_2023}.}
    \label{Xion_fesc_vs_EW_Ha}
\end{figure}

In Figure \ref{Xion_fesc_vs_EW_Ha}, we analyze the relation between $\mathrm{\xi_{ion,0}}$ and $\mathrm{EW_{0}(H\alpha)}$, which has already been estimated in \citet{Rinaldi_2023} for our HAEs. We notice that, among our sources, those that show both a high value of $\mathrm{EW_{0}(H\alpha)}$ and $\mathrm{\xi_{ion,0}}$ are also the youngest ones (see Table \ref{tab}). This result is consistent with what has been found in the recent literature, where young star-forming galaxies seem to show higher values of $\xi_{ion}$ \citep[e.g.,][]{Tang_2019}. 

By looking at Figure \ref{Xion_fesc_vs_EW_Ha}, we find quite a strong correlation between these two quantities, confirming what has been reported by \citet{Prieto_Lyon_2023} at $z \simeq 3-7$. We report data points from \citet{Harikane_2018, Lam_2019, Maseda_2020, Javi_2023} as well. In particular, the data point from \citet{Maseda_2020} seems to be off compared to our results, probably due to a much lower gas-phase metallicity that characterizes their sample \citep[see][for more details]{Maseda_2023}. The same trend has been reported also in \citet{Reddy_2018} for more massive galaxies at lower redshifts ($z\simeq1.4-3.8$). The correlation between $\xi_{ion,0}$ and $\mathrm{EW_{0}(H\alpha)}$, according to \citet{Tang_2019}, should hold only within the first 100 Myr since the onset
of star formation. Indeed, after 100 Myr, both young and intermediate-aged populations reach the equilibrium resulting, therefore, in a constant $L(H\alpha)$-to-$L(UV)$ ratio \citep{Atek_2022} and a plateau, at lower EWs, should arise in this comparison.

We also notice that the relation between $\xi_{ion,0}$ and EW$_{0}$(H$\alpha$) seems to saturate at very high EW$_{0}$ values, reaching a sort of plateau at EW$_{0}$ $\rm > 1000$ {\AA}. However, this claim must be taken with caution since a larger sample is needed to further constrain this result.

Finally, we want to highlight that the correlation between $\xi_{ion}$ and EW(H$\alpha$), as well as other nebular emission lines \citep[see][]{Prieto_Lyon_2023, Simmonds_2023} that are sensitive to ionization (e.g., [OIII]), can serve as a proxy for $\xi_{ion,0}$ at high redshifts, particularly when direct measurement of the rest-frame $L(UV)$ is not feasible.

\subsubsection{$\xi_{ion,0}$ versus specific Star Formation Rate}
\begin{figure}[ht!]
    \centering
    \includegraphics[width = 0.49 \textwidth, height = 0.30 \textheight]{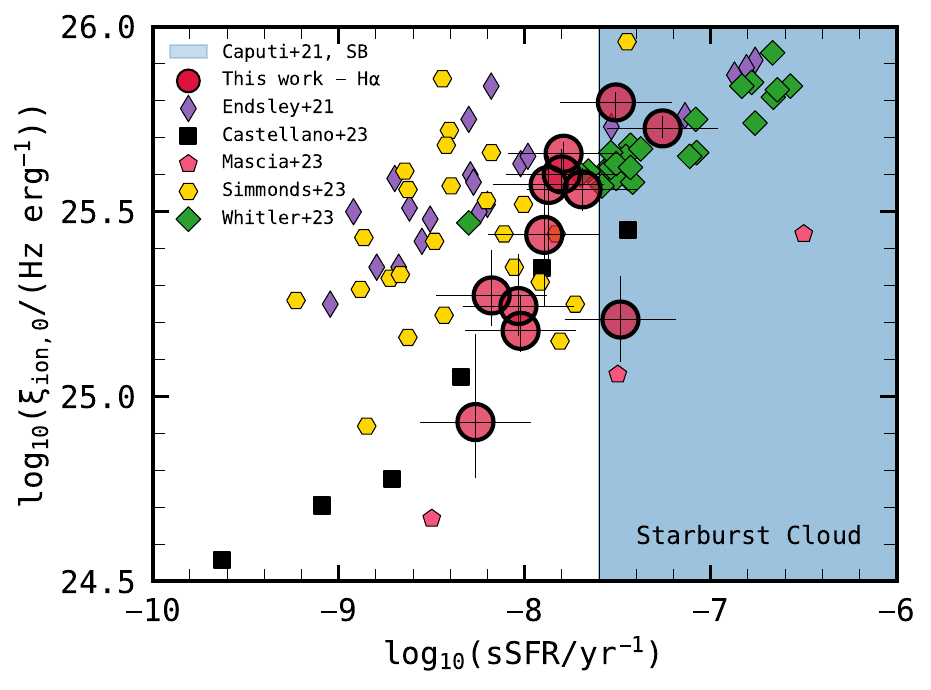}
    \caption{$\xi_{ion,0}$ versus sSFR. Recent findings from the literature are shown as well \citep{Endsley_2021, Castellano_2023, Mascia_2023b, Simmonds_2023, Whitler_2023}. The blue shaded area refers to the starburst region as defined in \citet{Caputi_2017, Caputi_2021}. In particular, for \citet{Castellano_2023} and \citet{Mascia_2023b} we show the median quantities. A strong correlation seems to arise from this comparison.}
    \label{Xion_vs_sSFR}
\end{figure}

We also investigate if there is any correlation between $\xi_{ion,0}$ and specific star formation rate (sSFR), which has been inferred from the H$\alpha$ emission line for our sample of HAEs \citep[]{Rinaldi_2023} -- see Figure \ref{Xion_vs_sSFR}. We collect data from the recent literature at high redshift as well. We find a positive correlation between those two parameters, where high values of sSFR correspond to high values of $\xi_{ion,0}$, as it has been reported at lower redshifts \citep[e.g.,][]{Castellano_2023, Izotov_2021}. In particular, this trend has been also suggested in \citet{Seeyave_2023} where they find, by exploiting \textit{First Light And Reionisation Epoch Simulations} (\textsc{Flares}), that $\xi_{ion}$ positively correlate with the sSFR. This finding probably indicates that galaxies that can double their stellar mass in a very short time (i.e., high sSFR) and, hence, are experiencing, at a fixed M$_{\star}$, a burst in terms of star formation can potentially produce a high fraction of ionizing photons that can escape the galaxy and, thus, reionize the surrounding medium. Interestingly, HAEs that happen to fall in the starburst cloud \citep[i.e., $\rm log_{10}(sSFR/yr^{-1}) \geq -7.60$;][]{Caputi_2017, Caputi_2021} are also among the youngest ones in our sample. Overall, by looking at the strong correlation between $\xi_{ion,0}$ and sSFR, this result may suggest that being young and starburst could have been crucial to producing a high fraction of ionizing photons. 

\subsubsection{$\xi_{ion,0}$ versus M$_{\star}$}
\begin{figure}[ht!]
    \centering
    \includegraphics[width = 0.49 \textwidth, height = 0.30 \textheight]{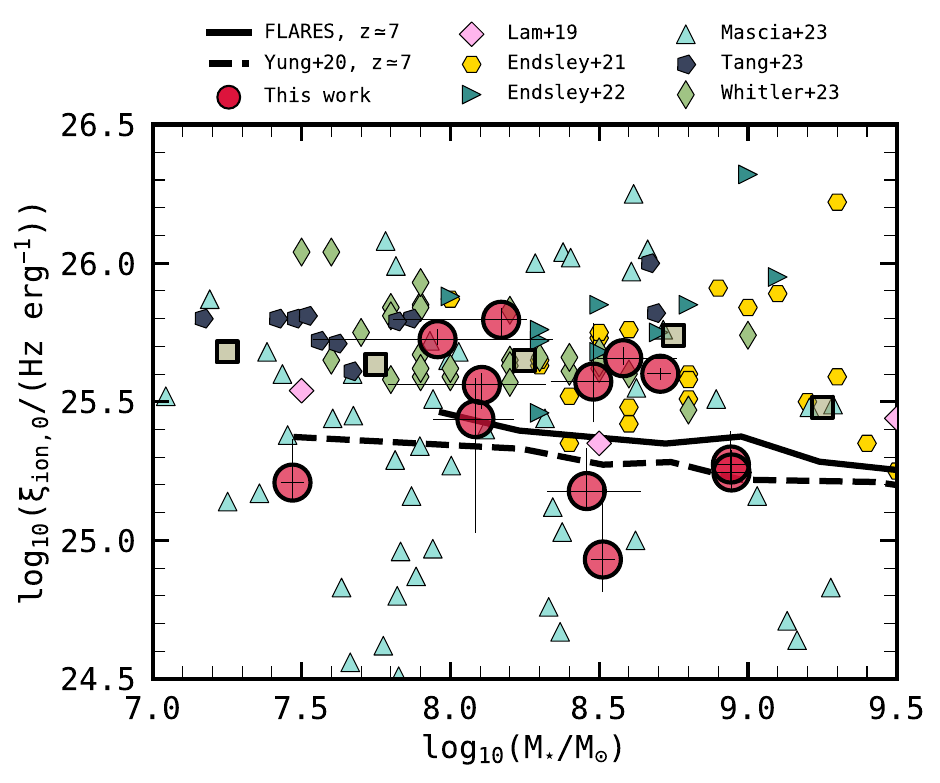}
    \caption{$\xi_{ion,0}$ versus M$_{\star}$. We also report recent findings at high redshift \citep{Lam_2019, Endsley_2021, Endsley_2022, Mascia_2023b, Tang_2023, Whitler_2023} as well as theoretical predictions from semi-analytical models \citep[dashed line,][]{Yung_2020} and hydrodynamical simulations \citep[solid line,][]{Seeyave_2023}, i.e. \textsc{Flares}. The square points refer to the median $\xi_{ion,0}$ per bin of stellar mass ($\rm \Delta M_{\star} = 0.5\;dex$). They show a weak anti-correlation, very similar to what is predicted from theoretical models.}
    \label{Xion_vs_mass}
\end{figure}

In Figure \ref{Xion_vs_mass}, we also study if there is any correlation between $\xi_{ion,0}$ and M$_{\star}$. To put everything in context, we collect data points from the most recent literature at high redshift as well. We find a weak anti-correlation between those two parameters, as shown by checking on Spearman's Rank correlation coefficient ($\rho \simeq -0.04$), where low-mass galaxies tend to have higher values of $\xi_{ion,0}$. Interestingly, the sample of low-mass galaxies we show in Figure \ref{Xion_vs_mass} (both our HAEs and galaxies from the literature) is characterized by having young ages. An anti-correlation between $\xi_{ion}$ and M$_{\star}$ has been also reported in \textsc{Flares} simulations \citep{Seeyave_2023} as well as by using semi-analytical models \citep[e.g.,][]{Yung_2020}, where they both conclude that low-mass galaxies could have been important contributors in Cosmic Reionization -- mostly because low-mass galaxies are more abundant than the massive ones, especially at high redshift \citep[e.g.,][]{Trebitsch_2022, Navarro_2023}. A similar trend has been reported at lower redshifts in \citet{Castellano_2023}, where they find a stronger anti-correlation than what we retrieve in our study -- mainly due to their larger sample. We also report, by adopting squares, the median trend of $\xi_{ion,0}$ as a function of M$_{\star}$ by binning galaxies in bins of stellar mass ($\Delta \rm M_{\star} = 0.5\;dex$). We recover the same trend as the simulations report but at slightly larger values of $\xi_{ion,0}$. A similar finding, by comparing simulations and observations, has been found in \citet{Seeyave_2023}.

\subsubsection{$\xi_{ion,0}$ versus UV-$\beta$ slope}
\begin{figure}[ht!]
    \centering
    \includegraphics[width = 0.49 \textwidth, height = 0.30 \textheight]{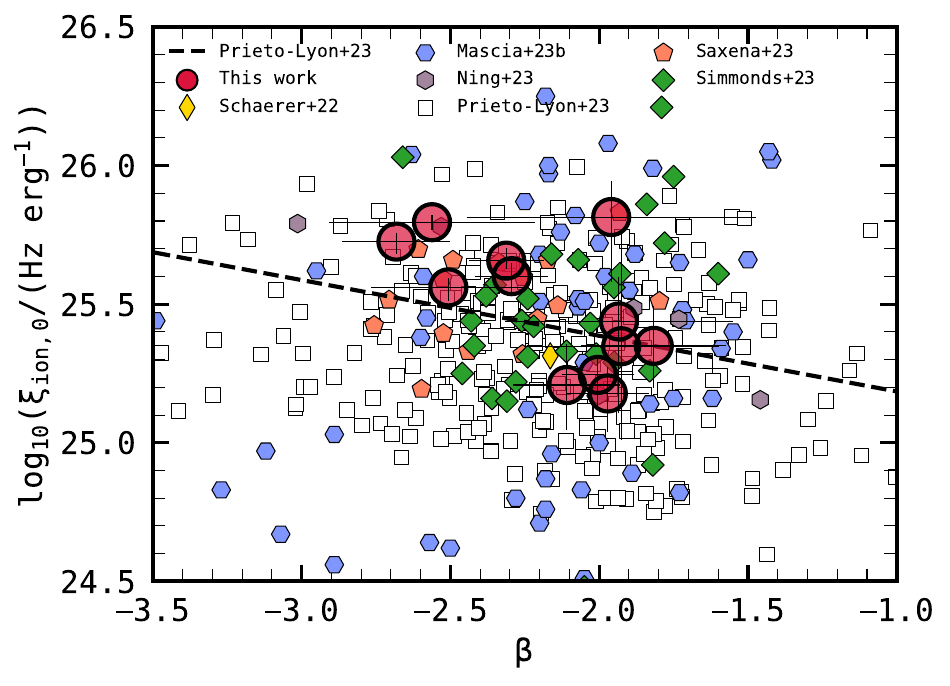}
    \caption{$\xi_{ion,0}$ as a function of $\beta$. We find a weak anti-correlation between $\beta$ and $\xi_{ion,0}$, as confirmed by checking on the Spearman's rank correlation coefficient. We report data points from literature at lower redshifts as well \citep{Schaerer_2022, Mascia_2023b, Ning_2023,  Prieto_Lyon_2023, Saxena_2023, Simmonds_2023}. The black dashed line refers to an anti-correlation between these two quantities that has been reported in \citep{Prieto_Lyon_2023}.}
    \label{Xion_fesc_vs_Beta}
\end{figure}

In Figure \ref{Xion_fesc_vs_Beta}, we analyze ${\xi_{ion,0}}$ as a function of the UV-$\beta$ slope. As we already mentioned above, the UV-$\beta$ slope is strictly related to both the metallicity and age of the stellar population (see Figure \ref{beta_vs_age_models}), therefore it can be related to the inferred ionization capability of a galaxy driven by its young stellar population \citep[e.g.,][]{Eldridge_2022}.

From Figure \ref{Xion_fesc_vs_Beta}, we see that there is a weak anti-correlation ($\rho \simeq -0.10$) between these two parameters, where $\xi_{ion}$ reaches the canonical value ($\rm log_{10} (\xi_{ion}/(Hz\;erg^{-1})) \simeq 25.2$) \sloppy at $\beta \simeq -2$ \citep[e.g.,][]{Robertson_2013} and shows an enhancement at $\beta < -2$. Recent observations have shown that galaxies at $z>6$, on average, have bluer UV-$\beta$ slopes compared to their low-z counterparts that could suggest an enhanced value of $\xi_{ion}$ at $z>6$. In particular, we can clearly see that our sample follows the same trend that has been reported in \citet{Prieto_Lyon_2023}, where they claimed a weak anti-correlation between $\mathrm{\xi_{ion,0}}$ and $\beta$. A similar trend has already been reported in the recent literature at $z\simeq6$ by making use of NIRCam data, where \citet[][]{Ning_2023} studied a sample of LAEs by analyzing their H$\alpha$ emission line.

\subsubsection{Inferring $f_{esc, LyC}$ from the UV-$\beta$ slope}
Since we can measure the UV-$\beta$ slopes for our galaxies, in fact, we can independently infer $f_{esc, LyC}$ following the prescription presented in \citet[][]{Chisholm_2022}. As we already mentioned before, estimating $f_{esc, LyC}$ at high redshifts is quite challenging. However, indirect indicators can be assumed to infer the escape fraction of Lyman continuum photons \citep[e.g.,][]{Chisholm_2022, Mascia_2023, Mascia_2023b}.

In this work, we make use of \citet{Chisholm_2022} results. They study low-redshift sources to investigate a possible correlation between $f_{esc, LyC}$, $\beta$, and M$_{UV}$ (see their paper for more details). We employ their derived prescription to infer $f_{esc, LyC}$ from their Equation (18):

\begin{equation}
{f_{esc, LyC} = (1.3\pm0.6)\times10^{-4}\times10^{(-1.2\pm0.1)\beta_{obs}}}.
\label{f_esc_equation}
\end{equation}

Interestingly, by considering our HAEs in terms of M$_{\star}$, $M_{UV}$, UV-$\beta$ slopes, and ages, we see that our sample resembles the parameter space presented in \citet{Chisholm_2022} (see their Figure 4, 9, and 11). This finding lends additional support to the method of employing Eq. \ref{f_esc_equation} for estimating the $f_{esc, LyC}$ for our sample of HAEs.

We find that most of the galaxies in our sample (75\%) show $f_{esc, LyC} \lesssim 10\%$. Only 25\% of our sample is characterized by a higher $f_{esc, LyC}$ value ($10\% \lesssim f_{esc, LyC} \lesssim 25\%$). In particular, our sample shows a median value of $f_{esc, LyC} \simeq 4\%^{+3}_{-2}$ (16th and 84th percentile), showing that the assumption $\xi_{ion} \simeq \xi_{ion,0}$ holds at these redshifts. Hereafter, for that reason, we will refer to $\xi_{ion}$ only in the subsequent figures.

Finally, here we do not compare $\xi_{ion}$ with the same properties as we did in the previous discussion because of the very low $f_{esc, LyC}$ values we retrieve from Equation \ref{f_esc_equation}. Indeed, the trends we find are very similar to what we already discussed above, therefore our conclusions do not change.

\subsection{The redshift evolution of $\xi_{ion}$}

\begin{figure*}[ht!]
    \centering
    \includegraphics{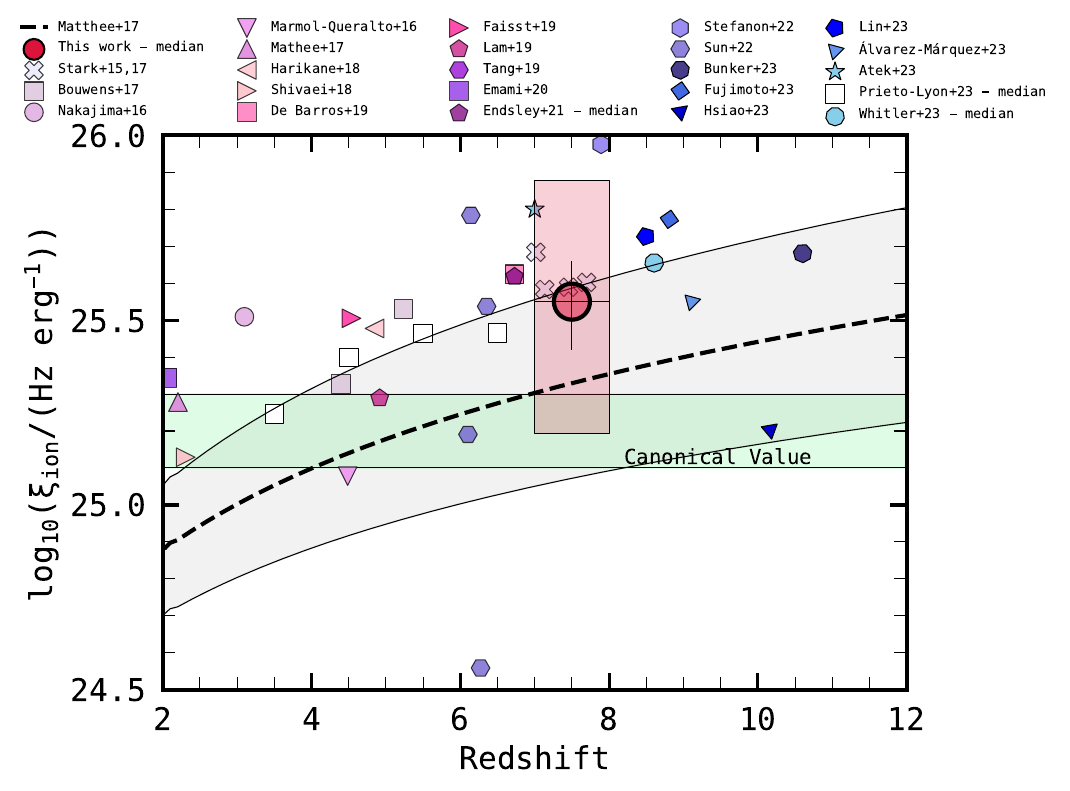}
    \caption{The evolution of $\xi_{ion}$ as a function of redshift. We report our results as well as a compilation of the recent literature at $z\simeq 1-12$ \citep{Stark_2015, Bouwens_2016, Marmol_2016, Nakajima_2016, Matthee_2017, Stark_2017, Harikane_2018, Shivaei_2018, De_Barros_2019, Faisst_2019, Lam_2019, Tang_2019, Emami_2020, Endsley_2021, Stefanon_2022, Sun_2022, Bunker_2023, Fujimoto_2023, Hsiao_2023, Lin_2023, Javi_2023, Prieto_Lyon_2023}. We find that $\xi_{ion}$ spans a large variety of values in our sample at $z\simeq 7-8$. The same variety of values has been found at lower redshifts \citep{Sun_2022, Ning_2023, Prieto_Lyon_2023}. We identify a mild evolution of $\xi_{ion}$ as a function of cosmic time.}
    \label{xion0_vs_redshift}
\end{figure*}

In Figure \ref{xion0_vs_redshift}, we show the redshift evolution of $\xi_{ion}$ in the context of the recent literature at $z\simeq1-12$ (see that plot for the references).

From this figure, we can notice that our sample spans a large variety of $\xi_{ion}$ values (red shaded area), showing a scatter that is similar to that already reported at both lower redshift \citep[e.g.,][]{Sun_2022, Prieto_Lyon_2023} as well as at higher redshift \citep[e.g.,][]{Whitler_2023}. This behaviour can be explained by taking into account the scattering due to the dust attenuation, different SFHs, and patchy ISM coverage \citep[e.g.,][]{Matthee_2017}. These results do not change even if we consider $\xi_{ion,0}$ (by assuming $f_{esc, LyC} \approx 0$ at high redshifts).  In particular, if we consider the median value of $\xi_{ion}$ at $z\simeq7-8$ we retrieve from our sample ($\rm log_{10}(\xi_{ion}/(\rm Hz\;erg^{-1})) = 25.55_{-0.13}^{+0.11}$), we find that it is in good agreement with the most recent results at similar redshifts \citep[e.g.,][]{Stefanon_2022, Sun_2022, Simmonds_2023}.

Furthermore, by considering our data points as well as those from the past literature, we can identify that there is a mild evolution of $\xi_{ion}$ as a function of redshift \citep[e.g., ][]{Matthee_2017, Stefanon_2022, Sun_2022} that can be explained by considering age effects: galaxies at higher redshifts have younger stellar populations and, therefore, higher $\xi_{ion}$ values. Nonetheless, metallicity effects could play a role as well. A similar result has been found from \citet{Atek_2023}, where they study a sample of 8 ultra-faint galaxies at $z\simeq 7$. Finally, in this work, we do not fit an evolution of $\xi_{ion}$ as a function of redshift due to the small size of our sample. Nevertheless, we notice that the evolution of $\xi_{ion}$ over cosmic time looks a bit steeper compared to what has been proposed in \citet{Matthee_2017}. However, a larger sample of galaxies at high redshift is needed to further constrain this result.

\section{Discussion: which sources drive reionization?}\label{section4}

\subsection{Implications for the escape fraction}

In this section, we evaluate the impact of ionizing production efficiency on the allowed escape fraction for our sample of HAEs at $z\simeq7-8$. As already mentioned before, in this work we find a slightly larger value of $\xi_{ion}$ ($\rm log_{10}(\xi_{ion}/(\rm Hz\;erg^{-1})) = 25.55_{-0.13}^{+0.11}$) compared to what has been previously found in the past \citep[e.g.,][]{Lam_2019} at lower redshifts.

\citet{Robertson_2013} showed that knowing $\xi_{ion}$ can help setting strong constraints on the escape fraction $f_{esc}$. Nevertheless, to do so, we need to make some assumptions. In particular, \citet{Robertson_2013, Robertson_2015} found an implicit constraint for $\xi_{ion}$ which is $\rm log_{10}(\xi_{ion}/(Hz\; erg\; s^{-1})) = 24.50 \pm 0.10$. Therefore, by following the same approach as \citet{Bouwens_2015, Lam_2019}, we can write a general formula for a wider range of faint-end cut-offs to the UV LF and clumping factors ($C$):

\begin{equation}
\begin{split}
   f_{\rm esc, rel} \xi_{\rm  ion} f_{\rm corr}(M_{\rm lim}) (C/3) ^{-0.3} = \\ 10^{24.50\pm0.10}\;\rm s^{-1}(erg\;s^{-1}\;Hz^{-1}),
    \label{f_esc_rel_equation}
\end{split}
\end{equation}
\noindent where $M_{\rm lim}$ is the UV luminosity cut-off and $f_{\rm corr}(M_{\rm lim})$ is a correction factor for $\rho_{UV} (z \simeq 7-8)$ \citep[see][for more details]{Bouwens_2015}. By looking at Equation \ref{f_esc_rel_equation}, we can clearly see that the product of $f_{\rm esc, rel} (\equiv f_{esc, LyC}/f_{esc,UV})$ and $\xi_{\rm ion}$ cannot be greater than what we retrieve from Equation \ref{f_esc_rel_equation} because, otherwise, the Cosmic Reionization should have been completed sooner compared to what we observe today \citep[$z \simeq 5-6$; e.g. ][]{Finkelstein_2019, Naidu_2020, Goto_2021, Bosman_2022}.

\begin{figure*}[ht!]
    \centering
    \includegraphics{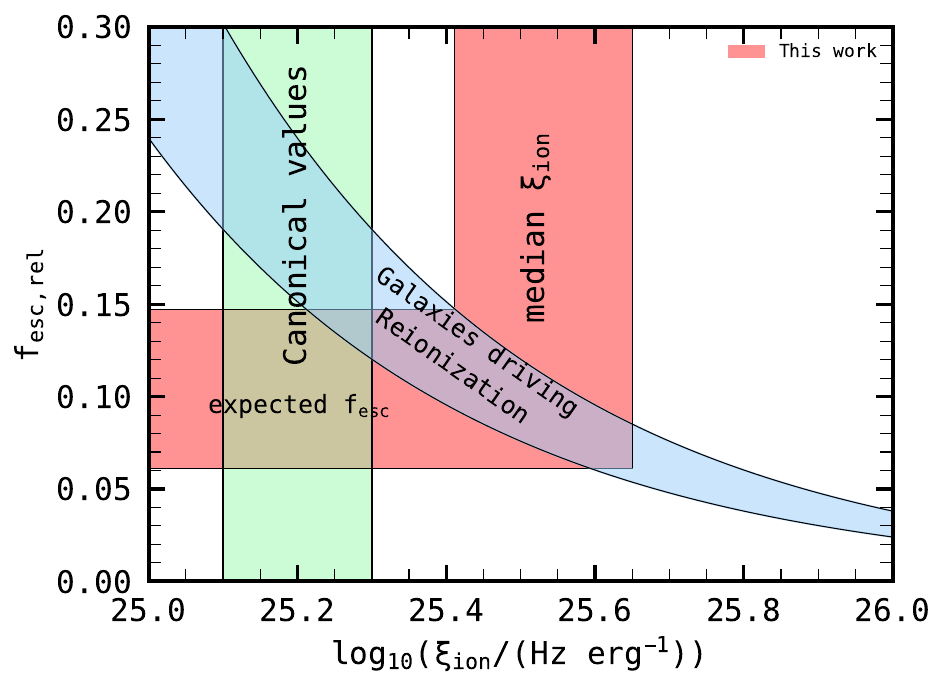}
    \caption{$f_{esc, rel} (\equiv f_{esc, LyC}/f_{esc, UV})$ as a function of $\xi_{ion}$. The green shaded area refers to the canonical value assumed for $\xi_{ion}$ \citep[e.g.,][]{Robertson_2013, Bouwens_2014}. The red shaded area refers to the $\xi_{ion}$ we inferred in this study. The blue shaded area has been derived by considering \citet{Bouwens_2015, Lam_2019} and assumptions from \citet{Robertson_2013, Robertson_2015}. The corresponding constraints we can place on the $f_{esc}$ ($6-15\%$) are indicated with a red shaded area.}
    \label{f_esc_rel_vs_xion0}
\end{figure*}

If we now assume that $M_{lim} = -13$ mag and $C = 3$, as proposed in the past literature \citep[e.g.,][]{Bolton_2007, Pawlik_2009, Shull_2012, Finlator_2012, Pawlik_2015}, from Figure \ref{f_esc_rel_vs_xion0} we find that $f_{esc, rel}$ does not need to be higher than $\simeq 6-15$ per cent for our sample of HAEs, at  $z\simeq7-8$ to have been able to reionize their surrounding medium. This finding seems to be in good agreement with what has been recently found in \citet{Atek_2023}, where they studied a sample of galaxies spectroscopically confirmed at high redshift ($z\simeq 7$) and conclude that galaxies might not have needed a large escape fraction of ionizing photons to reionize the surrounding medium.

Interestingly, our result is in line with what has been found in recent simulations like \textsc{SPHINIX} \citep{Rosdahl_2018} and \textsc{THESAN} \citep{Kannan_2022}. By looking at their simulations, they analyze the evolution of $f_{esc}$ as a function of the redshift. We find agreement between our result (from Figure \ref{f_esc_rel_vs_xion0}) and their theoretical predictions \citep{Rosdahl_2018, Yeh_2023} at $z\simeq7-8$. In particular, \citet{Yeh_2023}, from the \textsc{THESAN} simulations, studied $f_{esc}$ as a function of the redshift for different stellar masses, concluding that low-mass galaxies could have played an important role during Cosmic Reionization. \citet{Dayal_2020} found a similar result, by using semi-analytical models, where they found that the ionizing budget is dominated by stellar radiation from low-mass galaxies ($\lesssim 10^{9}\;\rm M_{\odot}$). Finally, a similar scenario has been also proposed by making use of observational constraints as well \citep{Meyer_2020, Davies_2021}.

\subsection{The ionizing emissivity of strong HAEs at $z\simeq 7-8$ and their role in Cosmic Reionization}

In this section, we investigate the possibility of our sample of HAEs driving Cosmic Reionization. In particular, we remind the reader that our sample of HAEs constitutes only $20\%$ of star-forming galaxies at $z\simeq 7-8$ \citep{Rinaldi_2023}. 

In evaluating the impact of these strong emitters in driving Cosmic Reionization, the total ionizing emissivity ($\dot{N}_{ion}$) constitutes a key ingredient. This quantity is typically estimated by considering three separate factors, assuming that galaxies produce the bulk of ionizing photons during Cosmic Reionization: the dust-corrected UV luminosity density ($\rho_{\rm UV}$), the ionizing photon production efficiency ($\xi_{ion}$), and the escape fraction of ionizing photons ($f_{esc}$):

\begin{equation}
\dot{N}_{ion} = \rho_{UV} \, \xi_{ion} \, f_{\rm esc}.
\end{equation}

To estimate $\rho_{UV}$, we integrate the UV Luminosity Function (LF) of our HAEs in the redshift bin studied in this work following the same apporach as outlined in \citet{Navarro_2023}. 

\begin{figure*}[ht!]
    \centering
    \includegraphics[width = 0.98 \textwidth, height = 0.60 \textheight]{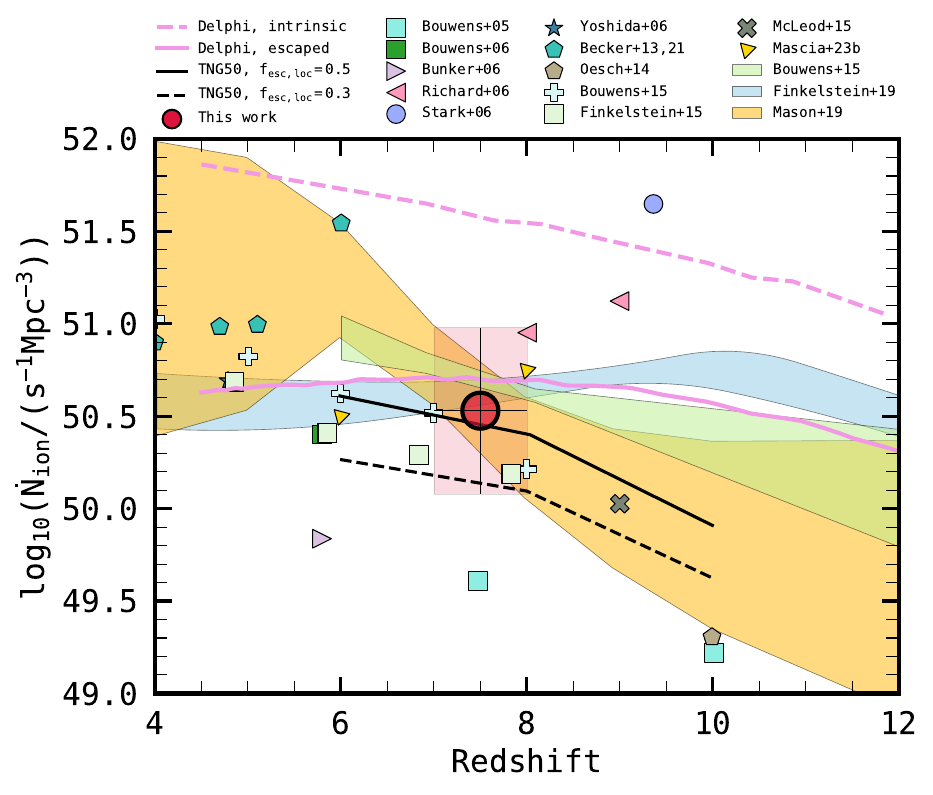}
    \caption{$\dot N_{ion}$ as a function of redshift, as obtained from our own data point and others from the literature  \citep{Bouwens_2005, Bouwens_2006, Bunker_2006, Richard_2006, Stark_2006, Yoshida_2006, Becker_2013, Oesch_2014, Bouwens_2015, Finkelstein_2015, McLeod_2015, Mascia_2023b}. A number of theoretical predictions from hydrodynamical and semi-analytical models \citep{Bouwens_2015, Finkelstein_2019, Manson_2019, Dayal_2022,  Kostyuk_2023} are also shown.}
    \label{N_emissivity_vs_redshift}
\end{figure*}

By considering $\rho_{UV}$, $\xi_{ion}$, and $f_{esc}$, we find that, at $z \simeq 7-8$, the expected total emissivity for our sample of HAEs should be $\dot N_{ion} = \rm 10^{50.53\pm0.45} \; s^{-1} Mpc^{-3}$, where the uncertainties on this quantity are mainly driven by the cosmic variance effect that affects our $\rho_{UV}$ estimate\footnote{{We adopted the same apporach as the one presented in \citet{Cosmic_Variance_2008} to take into account the cosmic variance;}}. We report this result in Figure \ref{N_emissivity_vs_redshift}. 

To evaluate the impact of strong HAEs during Cosmic Reionization, we considered the population of non-HAEs at $z\simeq7-8$ in our sample, with the latter representing 80\% of the total sample.

We remind the reader that the term “non-H$\alpha$ emitter” here refers to all galaxies except the ones identified as H$\alpha$ emitters in \citep{Rinaldi_2023}. This division is arbitrary and only given by MIRI's ability to detect the H$\alpha$ flux excess. The parameter $\xi_{ion}$, as well as all other parameters, most likely follow a continuum value distribution. However, analyzing the average properties of these two populations is still useful to compare how different these properties are between the most prominent line emitters and all other galaxies at similar redshifts.

In order to make a comparison between emitters and non-emitters at $z\simeq7-8$, for the non-emitters we assume an escape fraction\footnote{We considered the median value from our sample of HAEs as an upper limit for the non-HAEs, i.e. $f_{esc} = 7\%$;} and consider the canonical value for $\xi_{ion}$ that, for high-redshift sources, is $\rm log_{10}(\xi_{ion, 0}/(Hz;erg^{-1})) = 25.2$.
By making these assumptions, we find that $\dot N_{ion} = \rm 10^{50.10\pm0.45} \; s^{-1} Mpc^{-3}$ for the non-emitters, where HAEs contribute more than twice as much as non-HAEs within the same redshift bin. This result suggests that strong HAEs may have played an important role in terms of emitted ionizing photons per comoving volume at $z\simeq 7-8$. However, we wish to caution the reader that this conclusion is also contingent upon the assumed $\xi_{ion}$ for the non-HAEs at high-redshift, a parameter to which we lack direct access due to the absence of H$\alpha$ emission line detection.

In Figure \ref{N_emissivity_vs_redshift}, we show the contribution of our HAEs to the Cosmic Reionization by comparing our estimate of $\dot N_{ion}$ with the recent literature  in the context of its evolution over cosmic time. We compare our result to other observational constraints from \citet{Bouwens_2005, Bouwens_2006, Bunker_2006, Richard_2006, Stark_2006, Yoshida_2006, Becker_2013,  Oesch_2014, Bouwens_2015, Finkelstein_2015, McLeod_2015, Mascia_2023}. We also report theoretical models from \citet{Bouwens_2015, Finkelstein_2019, Manson_2019} as well as from \textsc{IllustrisTNG} simulations \citep{Kostyuk_2023}. In particular, we find that our result is in broad agreement with what has been previously reported in the literature at those redshifts. Finally, we report results from \textsc{Delphi} simulations that, however, predict a much higher value compared to our finding. We notice that the slight offset is due to the \textsc{Delphi} model including all the galaxies at $z\simeq7$ while we only consider strong line emitters.

\section{Conclusions} \label{section5}

In this paper, we analyzed a sample of H$\alpha$ emitters at $z\simeq 7-8$ that have been discovered in the Hubble eXtreme Deep Field thanks to the publicly available medium-band and broadband NIRCam imaging in the XDF, combined with the deepest MIRI 5.6$\mu m$ imaging existing in the same field \citep{Rinaldi_2023}.

The sample consists of the 12 most prominent HAEs at $z\simeq 7-8$, that account for $20\%$ of the star-forming galaxies at $z\simeq7-8$ \citep{Rinaldi_2023}.

By estimating their M$_{UV}$ and UV-$\beta$, we do not see any clear trend between these two parameters at $z\simeq7-8$, probably due to the fact that our sample is too small, although other studies, based on a much bigger sample, claimed its existence in the recent literature \citep[e.g.,][]{Cullen_2023}. (Figure \ref{beta_vs_MUV}).

By looking at our galaxies, we see that our HAEs have $\rm log_{10}(M_{\star}/M_{\odot}) \simeq 7.5 - 9$ and show a broad correlation between $\beta$ and M$_{\star}$ (Figure \ref{Beta_vs_Mass}). We notice that $\beta$ becomes bluer at lower M$_{\star}$, following the same results as shown in \citet{Finkelstein_2012b, Bhatawdekar_2021} at $z\simeq 7-8$. In particular, from Figure \ref{Beta_vs_Mass}, we notice that some of our very low-mass sources should be characterized by having a higher $f_{esc, LyC}$, as proposed in \citet{Chisholm_2022}.

Our sample of 12 HAEs at $z\simeq 7-8$ shows a large variety of UV-$\beta$ slopes (ranging from $\beta = -2.7$ to $\beta = -1.8$, with a median value of $\beta =-2.15 \pm 0.21$) as well as they are, on average, quite young ($\lesssim 35$ Myr), except for one single source that shows a stellar population a bit older compared to the rest of the sample ($\approx 300\;\rm Myr$) -- see Figure \ref{beta_vs_age_models}. 25\% of our sample shows very blue UV-$\beta$ slopes ($-2.7 \leq \beta \leq -2.5$), suggesting that they could be characterized by a large escape fraction of ionizing photons \citep{Chisholm_2022}.

Since we can estimate $L(H\alpha)$, our sample of HAEs allows us to estimate $\xi_{ion,0}$ ($\equiv \xi_{ion}$ when $f_{esc, LyC} = 0$, which is the common assumption at high redshifts). We find that our sources show a large variety of $\xi_{ion,0}$, with a median value of log$\mathrm{_{10}(\xi_{ion,0}/(Hz\;erg^{-1}))}$ $\simeq 25.50_{-0.12}^{+0.10}$. 

We then compared $\xi_{ion,0}$ with some other stellar properties we derived for this sample of HAEs.  A weak trend between $\xi_{ion,0}$ and M$_{UV}$ appears from Figure \ref{Xion_f_esc_MUV}, where, on average, fainter objects tend to have a slightly higher value of $\xi_{ion,0}$ -- also confirmed in the recent liteature \citep[e.g.,][]{Prieto_Lyon_2023, Simmonds_2023}.

We also studied if there is any relation between $\xi_{ion,0}$ and $\rm EW_{0}(H\alpha)$ (see Figure \ref{Xion_fesc_vs_EW_Ha}). We retrieve a correlation between these two quantities, as already pointed out in the literature \citep[e.g.,][]{Prieto_Lyon_2023, Ning_2023}. In particular, we find that, on average, galaxies with high $\xi_{ion,0}$ are the youngest ones and they tend to have higher sSFR (see Figure \ref{Xion_vs_sSFR}). We investigated if there was any relation between $\xi_{ion,0}$ and M$_{\star}$ (Figure \ref{Xion_vs_mass}). By comparing these quantities, we find a weak anti-correlation that suggests that low-mass galaxies are mainly characterized by having a larger value of $\xi_{ion,0}$, in agreement with what has been found at lower redshifts in \citet{Castellano_2023}. We also inspected if there was any trend between $\xi_{ion,0}$ and $\beta$. From Figure \ref{Xion_fesc_vs_Beta}, we find that there is a weak anti-correlation between those two quantities, which agrees with recent findings at lower redshifts \citep{Prieto_Lyon_2023}. In particular, galaxies with very blue UV-$\beta$ slopes tend to have a higher $\xi_{ion,0} (\equiv \xi_{ion}$ when $f_{esc, LyC} = 0)$. This behaviour can be linked to the fact that $\beta$ is strictly related to both the metallicity and age of the stellar population, as shown in Figure \ref{beta_vs_age_models}, and, thus, to the capability of a young stellar population to emit ionizing photons that can escape into the IGM.

By following some prescriptions presented in \citet{Chisholm_2022}, we inferred $f_{esc, LyC}$ (see Eq. \ref{f_esc_equation}). We find that most of our galaxies (75\%) show $f_{esc, LyC} \lesssim 10\%$. Only 25\% of our sample shows a higher $f_{esc, Lyc}$ ($10-20\%$). Since we inferred $f_{esc, LyC}$, we could estimate $\xi_{ion}$, that shows a median value of log$_{10}(\xi_{ion}/(\rm Hz\;erg^{-1})) = 25.55_{-0.13}^{+0.11}$. Since we find very low values of $f_{esc, LyC}$, with a median value of $4\%^{+3}_{-2}$, the aforementioned correlations and anti-correlations we found for $\xi_{ion,0}$ are still valid if we consider $\xi_{ion}$ instead.

We also investigated if there is an evolution of $\xi_{ion}$ as a function of the redshift (Figure \ref{xion0_vs_redshift}). We find that our sample spans a large variety of values of $\xi_{ion}$ at $z \simeq 7-8$, which is in line with the results both at lower redshifts \citep[e.g.,][]{Endsley_2021, Sun_2022, Prieto_Lyon_2023, Ning_2023} and higher redshifts \citep[e.g.,][]{Whitler_2023}. In this work, we cannot directly fit an evolution of this quantity as a function of the redshift given our sample size. We find that the median value of $\xi_{ion}$ we get from our sample is in agreement with the extrapolation at higher redshift of what has been proposed in \citet{Stefanon_2022, Sun_2022}. Moreover, we conclude that, on average, there is a mild evolution of $\xi_{ion}$ over cosmic time, as already suggested in the past \citep[e.g.,][]{Matthee_2017, Stefanon_2022, Sun_2022}, which looks a bit steeper than what has been proposed in the past \citep[][]{Matthee_2017}. However, a larger sample of galaxies at high redshift is needed to further constrain this finding.

Finally, we analyzed the role of our HAEs during Cosmic Reionization. To do so, we first estimate the maximum $f_{esc, rel}$ that our sources, assuming that star-forming galaxies drive the reionization, need to reionize the surrounding IGM. We find that it does not need to be higher than $6-15$ per cent, which is in agreement with what has been proposed in hydrodynamical simulations such as \textsc{SPHINIX} \citep{Rosdahl_2018} and \textsc{THESAN} \citep{Kannan_2022} where they study the evolution of the escape fraction over cosmic time and, in particular, focus on the role of low-mass galaxies in reionizing the Universe, suggesting that they could have played a key role. Then, we estimated the total ionizing emissivity $\dot{N}_{ion}$ as a function of redshift and put our results in the context of the recent literature. We find that $\dot{N}_{ion} = 10^{50.53 \pm 0.45}\rm \; s^{-1} Mpc^{-3}$ at $z\simeq 7-8$, which is more than twice as much as non-HAEs within the same redshift bin \citep{Rinaldi_2023}. We emphasize that our derived total ionizing emissivity corresponds only to the most prominent H$\alpha$ emitters ($\rm EW_{0}(H\alpha) \geq 239$ {\AA}, see \citealt{Rinaldi_2023}).

In light of our findings and in combination with what simulations predict, we can conclude that low-mass and young galaxies, undergoing an episode of star formation, could be potentially regarded as the primary agents for driving Cosmic Reionization. Particularly, by being strong H$\alpha$ emitters, this work suggests that these kind of sources may have potentially played a key role in terms of the number of ionizing photons injected in the surrounding IGM at $z\simeq 7-8$ and, for this reason, they need to be investigated more.  Deep \textit{JWST} observations are now showing us that we could potentially observe, more systematically, these strong emitters at high redshift giving us the unprecedented opportunity to finally constrain their role in Cosmic Reionization.

\acknowledgments
In memoriam to the MIRI European Consortium members Hans-Ulrik N\o{}rgaard-Nielsen and Olivier Le F\`evre.

The authors thank Maxime Trebitsch, Rafael Navarro-Carrera, and Paula Cáceres-Burgos for the useful discussions.
The authors thank Gonzalo Juan Prieto Lyon, Sara Mascia and Lily Whitler for providing their galaxy sample data in electronic format.
This work is based on observations made with the NASA/ESA/CSA James Webb Space Telescope. The data were obtained from the Mikulski Archive for Space Telescopes at the Space Telescope Science Institute, which is operated by the Association of Universities for Research in Astronomy, Inc., under NASA contract NAS 5-03127 for JWST. These observations are associated with programs GO \#1963, GO \#1895 and GTO \#1283. The authors acknowledge the team led by coPIs C. Williams, M. Maseda and S. Tacchella, and PI P. Oesch, for developing their respective observing programs with a zero-exclusive-access period. Also based on observations made with the NASA/ESA Hubble Space Telescope obtained from the Space Telescope Science Institute, which is operated by the Association of Universities for Research in Astronomy, Inc., under NASA contract NAS 5–26555. The specific observations analyzed can be accessed via:\dataset[DOI: 10.17909/gdyc-7g80, 10.17909/fsc4-dt61, 10.17909/fsc4-dt61, 10.17909/T91019, 10.17909/1rq3-8048, 10.17909/z2gw-mk31].
The work presented here is the effort of the entire MIRI team and the enthusiasm within the MIRI partnership is a significant factor in its success. MIRI draws on the scientific and technical expertise
of the following organisations: Ames Research Center, USA; Airbus Defence and Space, UK; CEA-Irfu, Saclay, France; Centre Spatial de Li\`{e}ge, Belgium;
Consejo Superior de Investigaciones Cient\'{\i}ficas, Spain; Carl Zeiss Optronics, Germany; Chalmers University of Technology, Sweden; Danish Space Research
Institute, Denmark; Dublin Institute for Advanced Studies, Ireland; European Space Agency, Netherlands; ETCA, Belgium; ETH Zurich, Switzerland; Goddard Space Flight Center, USA; Institute d’Astrophysique Spatiale, France; Instituto Nacional de T\'ecnica Aeroespacial, Spain; Institute for Astronomy, Edinburgh, UK; Jet Propulsion Laboratory, USA; Laboratoire d’Astrophysique de Marseille (LAM), France; Leiden University, Netherlands; Lockheed Advanced
Technology Center (USA); NOVA Opt-IR group at Dwingeloo, Netherlands; Northrop Grumman, USA; Max-Planck Institut für Astronomie (MPIA), Heidelberg, Germany; Laboratoire d’Etudes Spatiales et d’Instrumentation en Astrophysique (LESIA), France; Paul Scherrer Institut, Switzerland; Raytheon Vision Systems, USA; RUAG Aerospace, Switzerland; Rutherford Appleton Laboratory (RAL Space), UK; Space Telescope Science Institute, USA; Toegepast-
Natuurwetenschappelijk Onderzoek (TNO-TPD), Netherlands; UK Astronomy Technology Centre, UK; University College London, UK; University of Amsterdam, Netherlands; University of Arizona, USA; University of Cardiff, UK; University of Cologne, Germany; University of Ghent; University of Groningen, Netherlands; University of Leicester, UK; University of Leuven, Belgium; University of Stockholm, Sweden; Utah State University, USA.

KIC acknowledges funding from the Dutch Research Council (NWO) through the award of the Vici Grant VI.C.212.036. 
KIC and EI acknowledge funding from the Netherlands Research School for Astronomy (NOVA). The Cosmic Dawn Center is funded by the Danish National Research Foundation under grant No. 140. LC acknowledges financial support from Comunidad de Madrid under Atracci\'on de Talento grant 2018-T2/TIC-11612. SG acknowledges financial support from the Villum Young Investigator grant 37440 and 13160 and the Cosmic Dawn Center (DAWN), funded by the Danish National Research Foundation (DNRF) under grant No. 140. G.\"O., A.B. \&  J.M.  acknowledge support from the Swedish National Space Administration (SNSA). J.H. and D.L. were supported by a VILLUM FONDEN Investigator grant to J.H. (project number 16599).

JAM and ACG acknowledge support by grant PIB2021-127718NB-100 by the Spanish Ministry of Science and Innovation/State Agency of Research MCIN/AEI/10.13039/ 501100011033 and by “ERDF A way of making Europe”.

PGP-G acknowledges support from the Spanish Ministerio de Ciencia e Innovaci\'on MCIN/AEI/10.13039/501100011033 through grant PGC2018-093499-B-I00.

JPP and TVT acknowledge funding from the UK Science, Technology Facilities Council and the UK Space Agency.

PD acknowledges support from the NWO grant 016.VIDI.189.162 (“ODIN”) and from the European Commission's and University of Groningen's CO-FUND Rosalind Franklin program.

\begin{splitdeluxetable*}{l|cccccBccccc}

%\tabletypesize{\scriptsize}
\tablenum{1}
\tablecaption{The properties of H$\alpha$-emitters}
\tablewidth{0pt}
\tablehead{
\colhead{ID} & 
\colhead{R.A.}  & 
\colhead{Dec.}  & \colhead{$z_{phot}$}  & \colhead{log$_{10}$(Age/yr)} & \colhead{log$_{10}\rm(M_{\star}/M_{\odot})$} & \colhead{log$_{10}$(EW$\rm_{0}(H\alpha$)/{\AA})} & \colhead{$\beta$} & 
\colhead{M$_{UV}$} & \colhead{$f_{esc, LyC}$}&
\colhead{log$_{10}(\xi_{ion}/\rm Hz\;erg^{-1})$}}
\startdata
       MIDIS-7784 & 53.186448 & -27.779234 & 7.56 & 8.46$_{-0.74}^{+0.38}$ & $8.11_{-0.31}^{+0.21}$ & 2.94$_{-0.41}^{+0.34}$ & $-$2.51$_{-0.26}^{+0.26}$ & $-$18.51$_{-0.06}^{+0.06}$ & 0.15$_{-0.08}^{+0.24}$ & 25.63$_{-0.09}^{+0.15}$\\
  MIDIS-8868 & 53.176707 & -27.782018 & 6.98 & 6.34$_{-0.19}^{+0.36}$ & $8.48_{-0.14}^{+0.06}$ & 3.61$_{-0.08}^{+0.08}$ & $-$1.96$_{-0.48}^{+0.48}$ & $-$17.81$_{-0.16}^{+0.16}$ & 0.03$_{-0.01}^{+0.11}$ & 25.83$_{-0.23}^{+0.15}$\\
  MIDIS-9359 & 53.178683 & -27.776321 & 7.28 & 7.68$_{-0.25}^{+0.58}$ & $8.46_{-0.13}^{+0.18}$ & 2.72$_{-0.23}^{+0.20}$ & $-$1.97$_{-0.14}^{+0.14}$ & $-$18.89$_{-0.05}^{+0.05}$ & 0.03$_{-0.02}^{+0.03}$ & 25.19$_{-0.06}^{+0.10}$\\
  MIDIS-9432 & 53.179766 & -27.774649 & 7.20 & 6.50$_{-0.02}^{+0.19}$ & $8.94_{-0.07}^{+0.05}$ & 3.11$_{-0.12}^{+0.12}$ & $-$1.82$_{-0.24}^{+0.24}$ & $-$19.27$_{-0.08}^{+0.08}$ & 0.03$_{-0.02}^{+0.11}$ & 25.36$_{-0.23}^{+0.15}$\\
  MIDIS-9434 & 53.179546 & -27.774438 & 7.68 & 6.58$_{-0.43}^{+0.34}$ & $8.17_{-0.36}^{+0.09}$ & 3.56$_{-0.10}^{+0.10} $& $-$2.56$_{-0.35}^{+0.35}$ & $-$18.22$_{-0.06}^{+0.06}$ & 0.17$_{-0.10}^{+0.36}$ & 25.88$_{-0.08}^{+0.22}$\\
  MIDIS-9497 & 53.179550 & -27.773955 & 7.14 &  6.34$_{-0.43}^{+0.34}$ & $8.51_{-0.04}^{+0.04}$ & 3.08$_{-0.19}^{+0.18}$ & $-$1.93$_{-0.33}^{+0.33}$ & $-$17.92$_{-0.05}^{+0.05}$ & 0.03$_{-0.02}^{+0.06}$ & 25.36$_{-0.34}^{+0.43}$\\
  MIDIS-9553 & 53.179511 & -27.773457 & 7.58 & 6.34$_{-0.33}^{+0.64}$ & $8.08_{-0.14}^{+0.13}$ & 3.24$_{-0.37}^{+0.33}$ & $-$1.93$_{-0.38}^{+0.38}$ & $-$17.69$_{-0.08}^{+0.08} $ & 0.03$_{-0.02}^{+0.07}$ & 25.45$_{-0.19}^{+0.18}$\\
  MIDIS-9932 & 53.164649 & -27.788155 & 7.27 & 6.68$_{-0.29}^{+0.94}$ & $7.47_{-0.04}^{+0.04}$ & 2.82$_{-0.32}^{+0.27}$ & $-$2.11$_{-0.18}^{+0.18}$ & $-$17.88$_{-0.06}^{+0.06}$ & 0.05$_{-0.03}^{+0.06}$ & 25.23$_{-0.12}^{+0.13}$\\
  MIDIS-10026 & 53.164840 & -27.788268 & 7.16 & 6.48$_{-0.29}^{+0.61}$ & $8.71_{-0.04}^{+0.04}$ & 3.18$_{-0.11}^{+0.11}$ & $-$2.29$_{-0.12}^{+0.12}$ & $-$19.61$_{-0.04}^{+0.04}$ & 0.08$_{-0.04}^{+0.08}$ & 25.64$_{-0.07}^{+0.07}$\\
  MIDIS-10036 & 53.164696 & -27.788236 & 7.28 & 6.76$_{-0.20}^{+0.86} $& $8.58_{-0.09}^{+0.18} $& 2.89$_{-0.17}^{+0.15}$ & $-$2.31$_{-0.13}^{+0.13}$ & $-$19.33$_{-0.05}^{+0.05}$ & 0.09$_{-0.04}^{+0.08}$ & 25.70$_{-0.06}^{+0.06}$\\
  MIDIS-10874 & 53.161720 & -27.785397 & 7.31 & 6.58$_{-0.24}^{+0.13}$ & $8.89_{-0.08}^{+0.07}$ & 2.63$_{-0.27}^{+0.23}$ & $-$2.01$_{-0.12}^{+0.12}$ & $-$20.19$_{-0.05}^{+0.05}$ & 0.04$_{-0.02}^{+0.04}$ & 25.26$_{-0.09}^{+0.15}$\\
  MIDIS-13137 & 53.159856 & -27.770046 & 7.00 & 6.42$_{-0.47}^{+0.77}$ & $7.96_{-0.42}^{+0.39}$ & 3.60$_{-0.09}^{+0.09}$ & $-$2.68$_{-0.18}^{+0.18}$ & $-$18.73$_{-0.05}^{+0.05}$ & 0.24$_{-0.15}^{+0.37}$ & 25.85$_{-0.09}^{+0.27}$\\
\enddata
\tablecomments{We list the sample of 12 HAEs that have been selected in \citet[][]{Rinaldi_2023}. Redshifts, ages, and stellar masses have been obtained by running \textsc{LePHARE}. $\beta$ and M$_{UV}$ have been estimated by using the methodology explained in Section 3.1. $f_{esc, LyC}$ refers to the predicted escape fraction following the prescriptions presented in \citet{Chisholm_2022}. Finally, we report $\xi_{ion}$ (taking into account the predicted $f_{esc, LyC}$). }
\label{tab}

\end{splitdeluxetable*}

\vspace{5mm}
\facilities{{\sl HST}, {\sl JWST}}.

\software{\textsc{Astropy} \citep{astropy_2018}, 
          \textsc{LePHARE} \citep{LePhare_2011},
          \textsc{NumPy} \citep{Numpy},
          \textsc{pandas} \citep{Pandas}
          \textsc{Photutils} \citep{Photutils}, 
          \textsc{SciPy} \citep{Scipy}
          \textsc{SExtractor} \citep{SExtractor},
          \textsc{TOPCAT} \citep{Topcat}.
          }

\bibliography{References}{}
\bibliographystyle{aasjournal}
\end{document}